%% file: main.tex
\documentclass{article}
\usepackage{fancyhdr}

\usepackage{report}

\usepackage{subfigure}
\usepackage{soul}
\usepackage[utf8]{inputenc} 
\usepackage[T1]{fontenc}    
\usepackage{hyperref}       
\usepackage{url}            
\usepackage{booktabs}       
\usepackage{amsfonts}       
\usepackage{fancyhdr}       

\usepackage{nicefrac}       
\usepackage{microtype}      
\usepackage{graphicx}
\usepackage{pifont}
\usepackage{multirow}
\usepackage{CJKutf8}
\usepackage{listings}

\definecolor{darkmagenta}{rgb}{0.56, 0.0, 1.0}
\definecolor{softyellow}{rgb}{1.0, 0.92, 0.3} 
\definecolor{LightAquamarine}{rgb}{0.75, 1.0, 0.8} 
\definecolor{FireBrick}{RGB}{178,34,34}
\definecolor{MediumPurple}{RGB}{147,112,219}

\definecolor{uclablue}{rgb}{0.15, 0.45, 0.68}
\hypersetup{
    breaklinks,
    colorlinks=true,
    citecolor={darkmagenta},
    linkcolor={uclablue},
    urlcolor={uclablue}
}
\usepackage{wrapfig}
\usepackage{float}
\usepackage{subcaption}
\usepackage{placeins}
\usepackage{lipsum} 
\usepackage{tcolorbox}
\usepackage{amsmath}
\usepackage{amssymb}
\usepackage{utfsym}
\usepackage{fontawesome}
\usepackage{xspace}
\usepackage{enumitem}
\usepackage{multirow} 
\tcbuselibrary{breakable}
\usepackage{enumitem}
\usepackage{colortbl}
\usepackage{fancyhdr}

\usepackage{tcolorbox}
\usepackage{transparent}
\definecolor{lightgray}{gray}{0.9}
\definecolor{deepgreen}{RGB}{50,180,50}  
\definecolor{deepred}{RGB}{220,50,50}  

\definecolor{njuPurple}{RGB}{220,205,230}     
\definecolor{njuPurpleLight}{RGB}{250,245,252}   

\newtcolorbox{abstractbox}{
    colback=njuPurpleLight,   
    colframe=njuPurple,       
    boxrule=1pt,              
    arc=4mm,                  
    left=8pt,                 
    right=8pt,                
    top=8pt,                  
    bottom=8pt,               
    opacityback=0.95
}
\usepackage{pifont}      
\usepackage{tikz}

\usepackage{tcolorbox}
\tcbuselibrary{breakable}

\newtcolorbox{promptbox}[1]{
  colback=gray!5,
  colframe=gray!75,
  title=#1,
  fonttitle=\bfseries,
  breakable
}

\newcommand{\benchmark}{{SWE-Compass}}

\title{SWE-Compass: Towards Unified Evaluation of Agentic Coding Abilities for Large Language Models}

\author{
Jingxuan Xu$^{*}$, Ken Deng$^{*}$, Weihao Li$^{*}$, Songwei Yu$^{*}$, Huaixi Tang$^{*}$, Haoyang Huang$^{*}$, Zhiyi Lai$^{*}$, Zizheng Zhan$^{*}$, Yanan Wu$^{*}$, Chenchen Zhang$^{*}$, Kepeng Lei, Yifan Yao, Xinping Lei, Wenqiang Zhu, Zongxian Feng, Han Li, Junqi Xiong, Dailin Li, Zuchen Gao, Kun Wu, Wen Xiang, Ziqi Zhan, Yuanxing Zhang, Wuxuan Gong, Ziyuan Gao, Guanxiang Wang, Yirong Xue, Mengtong Li, Mengfei Xie, Xiaojiang Zhang, Jinghui Wang, Wenhao Zhuang, Zheng Lin, Huiming Wang, Zhaoxiang Zhang, Yuqun Zhang, Haotian Zhang, Bin Chen, Jiaheng Liu$^{\dagger}$
\\
\vspace{4mm}
\large
\textbf{Kuaishou Technology}, \textbf{Nanjing University}
\\
\vspace{2mm}
\texttt{xujingxuan05@kuaishou.com},    \texttt{dengken@kuaishou.com}, \texttt{liujiaheng@nju.edu.cn} \\
}

\begin{document}

\maketitle
\let\oldthefootnote\thefootnote

\let\thefootnote\relax\footnotetext{*~Equal Contribution. ~~$^\dagger$~Corresponding Author.}
\let\thefootnote\oldthefootnote

\begin{abstract}
Evaluating large language models (LLMs) for software engineering has been limited by narrow task coverage, language bias, and insufficient alignment with real-world developer workflows. Existing benchmarks often focus on algorithmic problems or Python-centric bug fixing, leaving critical dimensions of software engineering underexplored. To address these gaps, we introduce \textbf{SWE-Compass}\footnote{\url{https://huggingface.co/datasets/Kwaipilot/SWE-Compass}}, a comprehensive benchmark that unifies heterogeneous code-related evaluations into a structured and production-aligned framework. SWE-Compass spans 8 task types, 8 programming scenarios, and 10 programming languages,
with 2000 high-quality instances curated from authentic GitHub pull requests and refined through systematic filtering and validation.
We benchmark ten state-of-the-art LLMs under two agentic frameworks, SWE-Agent and Claude Code, revealing a clear hierarchy of difficulty across task types, languages, and scenarios.
Moreover, by aligning evaluation with real-world developer practices, we hope SWE-Compass can provide a rigorous and reproducible foundation for diagnosing and advancing agentic coding capabilities in large language models.

\end{abstract}

\section{Introduction}
Large language models (LLMs) trained on code have rapidly advanced from solving algorithmic puzzles to assisting with production-scale software development. Modern coding LLMs~\citep{kimiteam2025kimik2openagentic,meituanlongcatteam2025,anthropic2025systemcard,glm45agenticreasoningcoding,qwen3technicalreport} now exhibit strong multi-turn reasoning, long-context handling, and tool-use capabilities, enabling them to serve as autonomous coding agents that plan, edit, test, and deploy software. This shift has motivated a wave of benchmarks designed to measure their utility. However, existing evaluations fall short in capturing the full scope of real-world software engineering: most remain restricted to single-file tasks, Python-centric bug fixing, or synthetic algorithmic problems~\citep{codegeex,programsynthesislargelanguage,Li_2022,livecodeben,bigcodebench}, leaving critical developer activities such as feature implementation, refactoring, configuration, and performance optimization underexplored.

\begin{figure}[h]
  \hspace{-2.8cm}  
    \centering
    \begin{minipage}{0.495\textwidth}
        \centering
        \includegraphics[height=6cm]{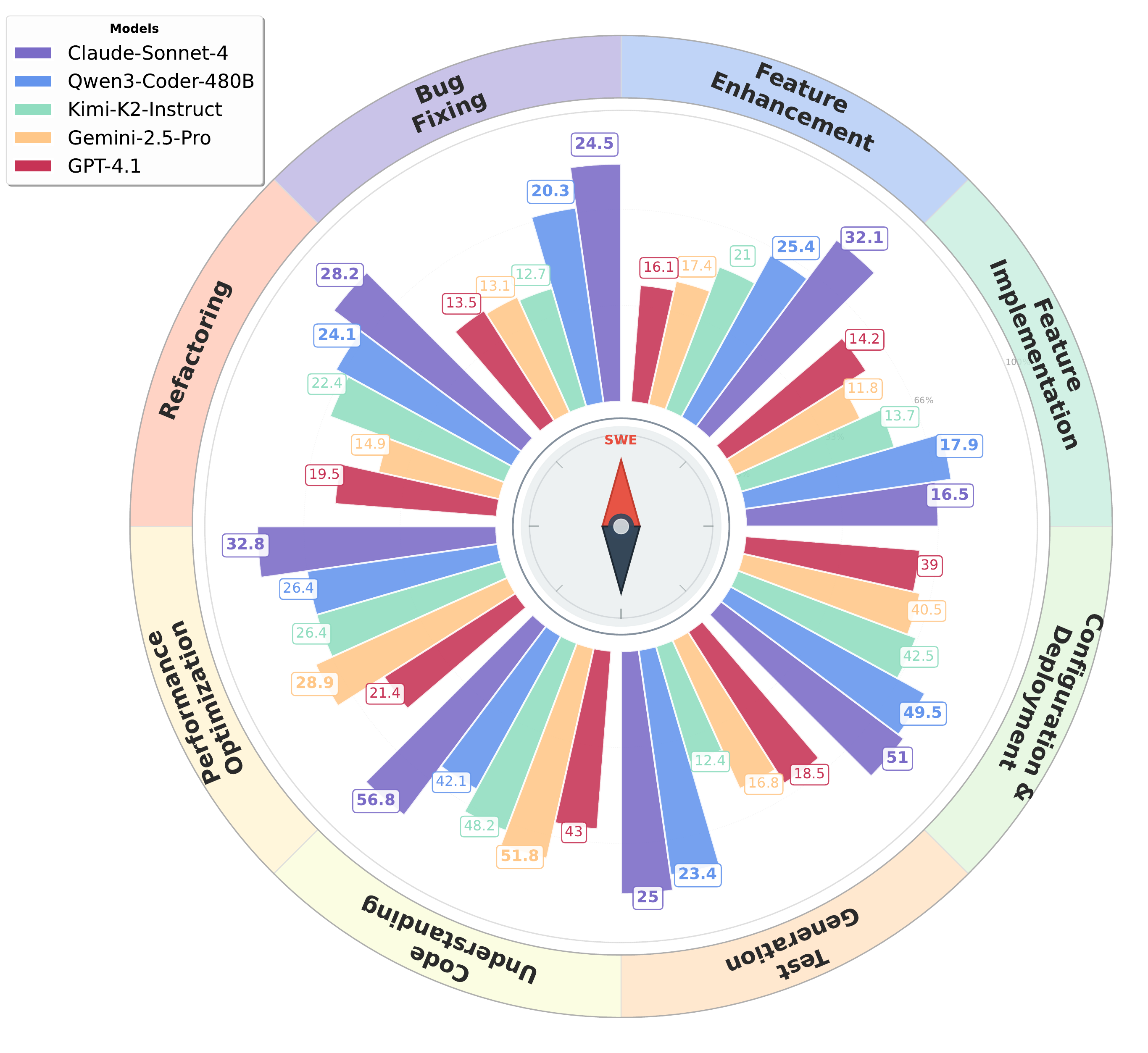}
        \captionof{subfigure}{Task-Specific Resolve Rates 
}
        \label{fig:task-dist}
    \end{minipage}%
    \hspace{-0.8cm}  
    \begin{minipage}{0.495\textwidth}
        \centering
        \includegraphics[height=6cm]{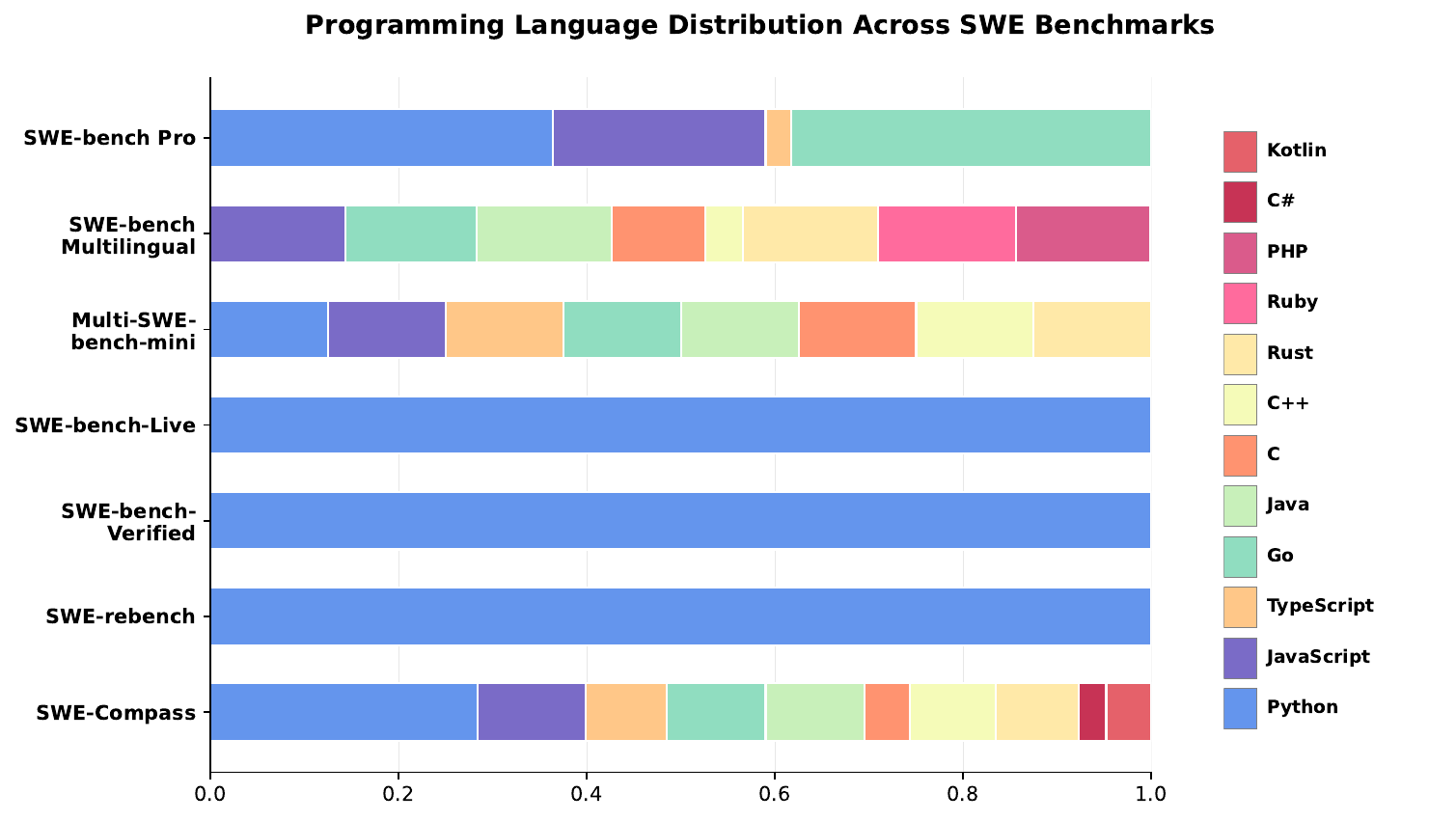}
        \captionof{subfigure}{Language Distribution Across Benchmarks}
        \label{fig:lang-scenario}
    \end{minipage}
    \caption{Comparative analysis: model performance across task types (left) and language coverage across benchmarks (right).}

    \label{fig:overview}
\end{figure}

Recent repository-grounded benchmarks, such as SWE-bench and its variants~\citep{jimenez2024swebench,yang2025swesmith,swerebench,yang2025swesmith}, have improved ecological validity by embedding evaluations in real issues, integrating test oracles, and introducing multi-language~\citep{swepolybench,yang2025swesmith} or multimodal~\citep{yang2025swebench} extensions. Yet these efforts largely converge on bug fixing as the dominant evaluation axis. As a result, they neglect the breadth of software engineering workflows that unfold across diverse scenarios—ranging from infrastructure and security engineering to machine learning system development—and across heterogeneous programming ecosystems. This narrowness prevents systematic capability diagnosis and obscures whether strong performance arises from generalizable reasoning or from artifact-specific adaptation.

To address these limitations, as shown in Figure~\ref{fig:overview}, we present \textbf{SWE-Compass}, a unified benchmark comprising 2,000 verified instances for evaluating LLMs' agentic coding abilities. SWE-Compass spans 8 task types, 10 programming scenarios, and 10 programming languages, combining broad coverage with rigorous evaluation fidelity. Each instance is paired with executable environments and reproducible tests, enabling fair comparison across prompting and agent-based methods under controlled budgets. Importantly, 
SWE-Compass is built upon four design principles: (i) real-world alignment, ensuring data originates from genuine developer interactions; (ii) comprehensive coverage across diverse tasks and languages; (iii) systematic taxonomy, providing structured labeling and balanced distributions; and (iv) evaluation fidelity, guaranteeing that all instances are executable and verifiable. Together, these principles yield a benchmark that reflects the complexity, diversity, and reproducibility demanded by modern software engineering.

Our contributions are threefold:

\begin{itemize}
    \item \textbf{A comprehensive, execution-grounded benchmark for software engineering.} We introduce SWE-Compass, a large-scale benchmark comprising 2,000 curated instances that span eight task types, eight programming scenarios, and ten programming languages. Each instance is drawn from real-world GitHub pull requests and paired with a reproducible execution environment, enabling rigorous and faithful evaluation of model performance in realistic development workflows.
    \item \textbf{A systematic evaluation framework aligned with real-world developer activities.} 
    SWE-Compass establishes a structured taxonomy to assess models across different dimensions such as feature implementation, refactoring, test generation, and deployment. This design enables fine-grained diagnosis of LLM capabilities and provides a principled foundation for comparing agentic coding systems under consistent conditions.
    %
    \item \textbf{Comprehensive empirical analysis and insights into LLM coding behavior.} Experiments with state-of-the-art LLMs and agentic systems reveal persistent gaps across tasks, languages, and scenarios, highlighting the difficulty of scaling beyond bug fixing and emphasizing the need for benchmarks that reflect the full complexity of real-world software engineering. 

\end{itemize}

\section{Related Works}

\noindent\textbf{Coding LLMs and Agents.} Code large language models (Code LLMs)~\citep{Chen2021Evaluating, Zhao2024CodeGemmaOC, chowdhery2023palm, nijkamp2022codegen, fried2022incoder, xu2022systematic, codellama, qwen25coder, deng2025hipo,que2024d} — excel at a wide range of programming tasks, including code generation, completion, repair, translation, code comprehension, documentation generation, and cross-language migration, among others. Crucially, modern Code LLMs combine ultra-long context support with robust tool-calling capabilities, enabling them to maintain global awareness across large codebases while actively invoking editors, shells, debuggers, or web browsers~\citep{liu2024ddk,liu2025comprehensive,wang2024mtu}. This synergy has fueled the rise of agentic coding systems — such as SWE-Agents~\citep{sweagent}, OpenHands~\citep{wang2025openhands}, and Claude Code~\citep{ClaudeCode}, QwenCode~\citep{qwen3technicalreport}, Codex~\citep{openai_codex}, Cline~\citep{Cline} — that autonomously plan, search, edit, test, and even perform agentic browser use to fetch live API documentation or solutions. As evaluations move toward dynamic, repository-scale workflows, these agent-based systems are showing improved performance over traditional code-generation approaches — particularly in tasks requiring persistent context, environment interaction, and multi-step reasoning~\citep{liu2024roleagent,he2025can}.

\noindent\textbf{Coding Benchmarks.} Single-file code benchmarks — such as HumanEval~\citep{codegeex}, MBPP~\citep{programsynthesislargelanguage}, CodeContests~\citep{Li_2022}, LiveCodeBench~\citep{livecodeben} and BigCodeBench~\citep{bigcodebench} — evaluate models on isolated algorithmic problems, abstracting away the structural, contextual, and environmental complexity inherent in real-world software engineering~\citep{liu2024m2rc}; while SWE-bench~\citep{jimenez2024swebench} and its variants — including Multimodal SWE-bench~\citep{yang2024swebenchmultimodal}, SWE-bench Multilingual~\citep{yang2025swesmith}, SWE-bench-Live~\citep{zhang2025swebenchgoeslive}, SWE-Lancer~\citep{swelancer},SWE-rebench~\citep{swerebench} and others — have substantially improved ecological validity by grounding evaluation in real repository issues and incorporating dimensions such as visual context, multi-language support, tool interaction, and repository-scale execution, they remain overwhelmingly confined to bug fixing as the de facto evaluation paradigm, neglecting the broader spectrum of developer activities such as feature implementation, refactoring, configuration, performance optimization, and test generation, which unfold across diverse engineering contexts including application and infrastructure development, ML/AI systems, security, UI/UX, and beyond — a critical omission that precludes fine-grained, scenario-aware capability analysis and obscures whether model performance stems from general reasoning, domain adaptation, or artifact overfitting; to address this gap, we introduce a benchmark that explicitly structures evaluation along orthogonal axes of task type and programming scenario, enabling systematic diagnosis of model strengths and weaknesses across the multifaceted reality of software development, rather than reducing it to a single, narrow slice.

\input{sec/3_method.tex}

\input{sec/4_experiments.tex}

\bibliographystyle{unsrtnat}
\bibliography{main} 
\input{sec/appendix.tex}

\end{document}

%% file: sec/3_method.tex



\section{SWE-Compass}
\label{sec:method}

\subsection{Overview}

\begin{figure}[t]
  \centering
  \includegraphics[width=1.0\linewidth]{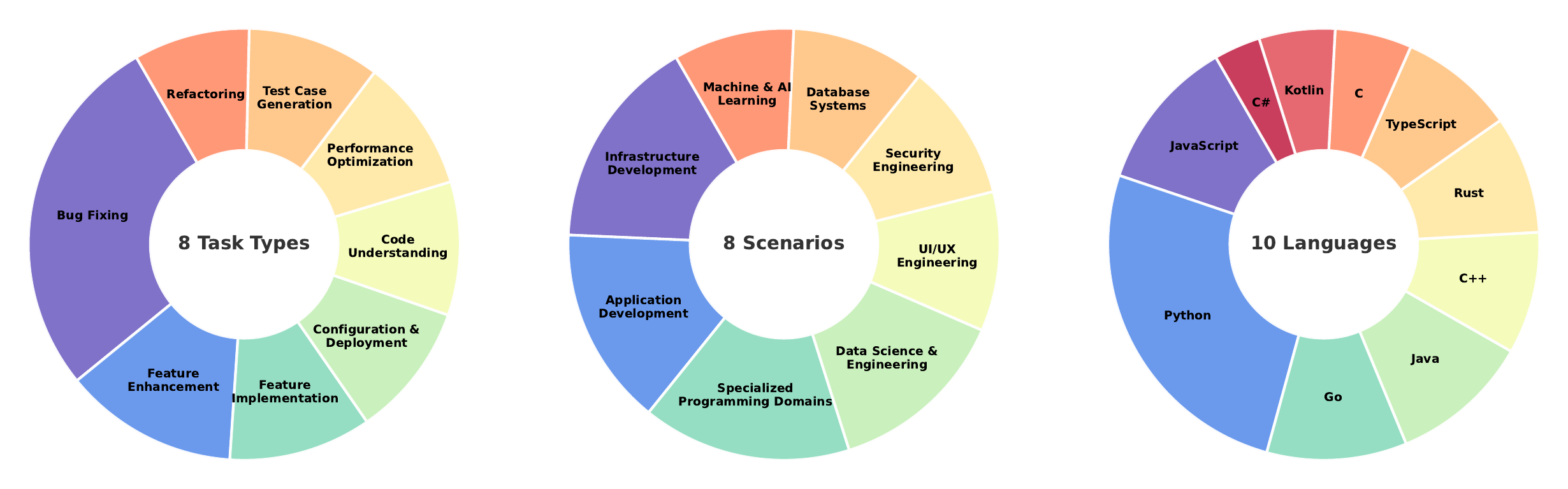}
  \caption{Distributions across task types, programming scenarios and languages.}
  \label{fig:dist_pies}
\end{figure}

Existing software engineering benchmarks primarily focus on Python-centric bug fixing tasks, exhibiting limited task coverage and insufficient alignment with real-world developer activities. In contrast to such benchmarks that concentrate on a single programming language and task type, SWE-Compass is constructed from authentic software engineering requirements, as shown in Table \ref{tab:benchmark_comparison}. It collects a large volume of high-quality repositories from GitHub pull requests and undergoes a multi-stage filtering and construction process. The resulting benchmark encompasses 2000 instances across \textbf{8 types of code-related tasks}, \textbf{8 programming scenarios}, and \textbf{10 programming languages}, as shown in Figure \ref{fig:dist_pies}. It enables a comprehensive evaluation of key software engineering capabilities, including bug fixing, performance optimization, and other related tasks, offering a holistic assessment of model performance in realistic software engineering contexts.

\begin{table}[t]
\caption{Comprehensive comparison of SWE-Compass with existing benchmarks across different dimensions.}
\label{tab:benchmark_comparison}

\centering
\small
\resizebox{0.99\textwidth}{!}{
\begin{tabular}{lccccc}
\toprule
\textbf{Benchmark} & \textbf{\# Samples} & \textbf{Languages}  & \textbf{Task Types} & \textbf{\# Repos} & \textbf{\# Mod. Files (Avg.)}\\
\midrule
HumanEval & 164 & Python  & Algorithm & --- & 1.0 \\
SWE-Bench-Verified & 500 & Python & Bug Fixing & 12 & 1.3 \\
SWE-Bench-Live & 1,319 & Python  & Bug Fixing & 93 & 3.3 \\
SWE-Bench-Multilingual & 300 & 9  & Bug Fixing & 42 & 1.3 \\
Multi-SWE-Bench & 1,632 & 7 & Bug Fixing  & 39 & 4.9 \\
SWE-Bench-Pro & 1,865 & 4 & 4 Types & 41 & 4.1 \\
\midrule
\textbf{SWE-Compass (Ours)} & \textbf{2,000} & \textbf{10}  & \textbf{8 Types} & \textbf{40} & \textbf{4.7} \\
\bottomrule
\end{tabular}}
\end{table}

\subsection{Design Principles}
SWE-Compass is designed around four guiding principles that distinguish it from existing software engineering benchmarks:
\begin{itemize}[left=10pt] 
   \item \textbf{Real-World Alignment}: The benchmark is grounded in authentic developer workflows by collecting tasks from large-scale discussions on GitHub and Stack Overflow. This ensures that evaluation scenarios directly reflect the diversity and complexity of real-world software engineering requirements rather than simplified or synthetic problem settings.
  \item \textbf{Comprehensive and Balanced Coverage}: SWE-Compass systematically spans the full spectrum of software engineering activities—including implementation, enhancement, maintenance, testing, and deployment—while ensuring balanced distributions across diverse programming scenarios and over 10 programming languages. Unlike prior benchmarks, which are heavily skewed toward Python-centric bug fixing tasks, SWE-Compass deliberately broadens coverage to underrepresented categories such as refactoring, performance optimization, and code understanding. 
  \item \textbf{Systematic Taxonomy}: Through an iterative active learning pipeline, SWE-Compass distills raw developer discussions into a structured taxonomy of task types, programming scenarios, and languages. This taxonomy provides a principled framework for data collection, classification, and synthesis, ensuring both granularity and scalability.
  \item \textbf{Evaluation Fidelity}: All benchmark instances are tied to executable test patches and reproducible environments, ensuring that evaluation results reflect genuine functional correctness. For task categories underrepresented in real-world repositories, SWE-Compass supplements data through carefully controlled synthesis while maintaining consistency with real-world task characteristics.
\end{itemize}
\begin{figure}
    \centering
    \includegraphics[width=1.0\linewidth]{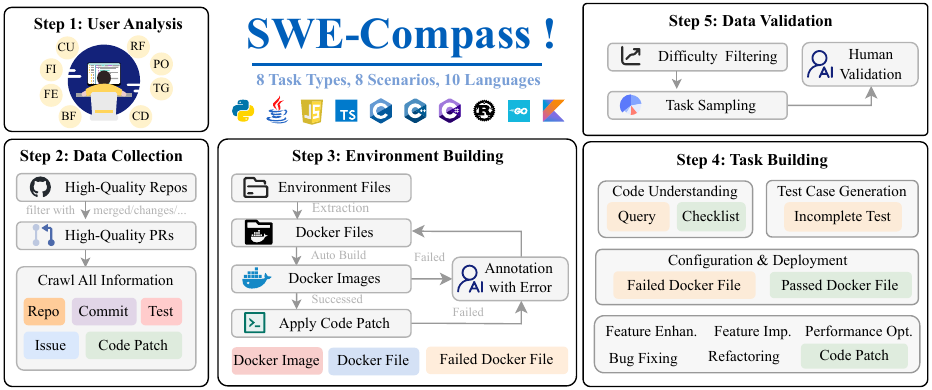}
    \caption{Construction of SWE-Compass.}
    \label{fig:method}
\end{figure}

\subsection{Benchmark Construction}
The construction of SWE-Compass follows a systematic and scalable approach organized into five major steps to ensure comprehensive coverage, balance, and real-world relevance: \textbf{(1) user analysis}, \textbf{(2) data collection}, \textbf{(3) environment building}, \textbf{(4) task construction}, and \textbf{(5) data validation}, as illustrated in Figure \ref{fig:method}. Specifically, through an iterative \textit{Active Learning} procedure applied to real-world coding conversations, we first identified that user needs predominantly fall into eight distinct task types, eight representative programming scenarios, and ten programming languages. We then collected a large volume of high-quality pull request (PR) data from GitHub repositories. By combining automated processing with expert annotation, we successfully built a set of executable development environments. Next, for each of the eight task types, we constructed and synthesized the corresponding task instances. Finally, after a multi-round filtering and quality validation process, we curated the SWE-Compass benchmark as the final dataset.


\subsubsection{Step 1: User Analysis}
To ensure that the evaluation accurately reflects model capabilities in realistic software development contexts, we collected repository-level coding discussions from two major platforms—\textbf{Stack Overflow} and \textbf{GitHub}. To discover emerging task categories, we designed an automated Active Learning framework for category discovery. Specifically, four popular software-related topics were chosen as initial label seeds for both task types and programming scenarios. Using an In-Context Learning (ICL)-based labeling approach, a large language model (LLM) was employed to annotate the collected conversations across three dimensions: task type, programming scenario, and programming language. Subsequently, tag clustering and LLM-guided seed optimization (via addition, modification, or deletion of tags) were applied to refine the label pool. The iterative process continued until convergence—when the updated seed pool no longer significantly differed from the previous ICL-generated pool. In our experiments, the {\texttt{Qwen3-Coder-30B-A3B-Instruct}}\footnote{https://huggingface.co/Qwen/Qwen3-Coder-30B-A3B-Instruct} model was used as the LLM annotator, and five iterations were performed in total. Ultimately, we identified \textbf{eight task types}, \textbf{eight programming scenarios}, and \textbf{ten major programming languages} as follows.

\noindent \underline{Task Types: }
\begin{itemize}[left=10pt]
\item \textit{Feature Implementation (FI)}: Developing features or modules from scratch, representing a core activity distinct from modifications or bug fixes.
\item \textit{Feature Enhancement (FE)}: Modifying or enhancing existing features to improve functionality, excluding any bug-related changes.
\item \textit{Bug Fixing (BF)}: Identifying, diagnosing, and resolving defects in the code, including troubleshooting and debugging.
\item \textit{Refactoring (RF)}: Improving the structure and maintainability of the code without altering its external behavior or functionality.
\item \textit{Performance Optimization (PO)}: Enhancing system efficiency and resource utilization, focusing specifically on performance improvements and distinct from refactoring.
\item \textit{Code Understanding (CU)}: Exploring, analyzing, and understanding code through static and dynamic analysis, including generating reports.
\item \textit{Test Case Generation (TG)}: Automatically generating unit and integration tests to validate code and ensure quality assurance.
\item \textit{Configuration \& Deployment (CD)}: Setting up environments, managing dependencies, and writing deployment scripts to ensure smooth application operation.
\end{itemize}

\noindent \underline{Programming Scenarios:}
\begin{itemize}[left=10pt]
\item \textit{Application Development (AD)}: Developing applications for specific environments such as web or desktop platforms, with an emphasis on feature implementation and platform adaptation.
\item \textit{Database Systems (DS)}: Designing, developing, managing, and optimizing databases to ensure efficient data storage, access, and consistency.
\item \textit{Data Science \& Engineering (DE)}: Handling data processing, analysis, mining, ETL, and feature engineering, emphasizing data-driven decision-making and efficient pipeline construction.
\item \textit{Machine Learning \& AI (ML)}: Training models, building recommendation systems, applying algorithms to enable intelligent decision-making and predictions.
\item \textit{Infrastructure Development (ID)}: Building foundational systems such as distributed architectures, system deployment, and DevOps tools, emphasizing stability, scalability, and automation.
\item \textit{Specialized Programming Domains (SPD)}: Addressing areas such as graphics, gaming, multimedia, and networking that require specialized technical expertise and tailored solutions.
\item \textit{Security Engineering (SE)}: Ensuring application and system security, identifying vulnerabilities, and implementing measures such as encryption to maintain compliance with security standards.
\item \textit{UI/UX Engineering (UI/UX)}: Designing and optimizing user interfaces and experiences across platforms to enhance visual appeal, usability, and consistency.
\end{itemize}

\noindent \underline{Programming Languages: } Python, JavaScript, TypeScript, Java, C, C++, Go, Rust, Kotlin, C\#.

\subsubsection{Step 2: Data Collection}
To ensure both coverage and realism in benchmark construction, we conducted the following data collection process.
Specifically, we first gathered existing open-source SWE benchmarks (e.g., Python bug fixing datasets) and mapped them to our defined taxonomy of task types, programming scenarios, and languages. 
As shown in Appendix \ref{opensource_dis_limit}, these benchmarks exhibit severe deficiencies across multiple dimensions: many task types are missing, scenario distributions are highly imbalanced, and programming languages are heavily skewed. 
To address these limitations, we further supplemented the dataset with high-quality repositories from GitHub in the following strategy.
\begin{itemize}[left=10pt]
\item \textbf{{High-Quality Repository Acquisition.}} Repositories were filtered using multiple quality indicators, including: valid open-source licenses, at least 500 stars, active maintenance within the past six months, at least three distinct contributors, more than 1000 issues and PRs, more than 200 forks, and the presence of executable unit tests.
This process yielded a large set of diverse, actively maintained repositories across 10 programming languages.
\item \textbf{{High-Quality PR Acquisition.}} Within the filtered repositories, we extracted all associated PRs and applied multi-stage filtering to retain only those with clear and meaningful modification semantics.
Specifically, we kept PRs that were successfully merged into the main branch, linked to descriptive Issues, and contained identifiable file- or line-level changes.
Each retained PR was required to have complete metadata, including repository, issue description, commit, test patch, and code patch.
\end{itemize}
After all filtering stages, approximately \textbf{50,000 high-quality PRs} were preserved, serving as the foundation for subsequent environment and task construction.

\subsubsection{Step 3: Environment Building} 
To enable reproducible execution and evaluation of each software engineering instance, we constructed isolated containerized environments for all selected PRs.
For each PR, we automatically extracted environment dependency information—such as package managers, required libraries, build tools, and runtime versions—from configuration files (e.g., \texttt{requirements.txt}, \texttt{setup.py}, \texttt{Makefile}, and CI/CD scripts). These dependencies were programmatically organized into corresponding Dockerfiles, from which initial Docker images were generated. 

Each successfully built image was then validated by executing the repository’s native test suite to verify that it could run end-to-end and reproduce the functionality and performance behavior before and after patch application (F2P/P2P consistency). The initial automated build success rate was around 2\%, reflecting the inherent complexity and dependency fragility in real-world repositories.

To address build failures, 30 expert annotators inspected the corresponding build logs, identified root causes (e.g., missing dependencies, version conflicts, or OS-level mismatches), and applied targeted fixes before re-triggering the build on Kubernetes. This expert-assisted retry process raised the overall retention rate to approximately 8\%.

Finally, we obtained about \textbf{4,000 successfully runnable Docker images}, each providing a fully reproducible and verifiable execution environment for downstream task synthesis and evaluation.

\subsubsection{Step 4: Task Building}
Given the heterogeneity of the eight software engineering task types, we designed three complementary strategies to construct diverse and representative task instances: \textbf{(1) Checklist Synthesis}, \textbf{(2) Reverse Masking}, and \textbf{(3) Targeted Filtering}. Each strategy was tailored to the specific characteristics of the corresponding task type, ensuring both task realism and evaluation reliability.

\begin{itemize}[left=10pt]
\item \textbf{{Checklist Synthesis.}}  For the \textbf{\textit{Code Understanding}} task, where the goal is to evaluate a model’s capability to comprehend and reason about code semantics, we adopted a data synthesis generation pipeline. 
Each instance was built from a combination of \texttt{Issue}, \texttt{Code Patch}, and \texttt{Test Patch} extracted from real PRs. 
We then prompted GPT-5 to generate multiple natural-language \textbf{queries} for each PR (e.g., \textit{Which functionality does this change affect?}). 
These generated queries were filtered by a difficulty-aware scoring function to remove trivial or ambiguous cases. 
For each remaining query, GPT-5 was further instructed to produce \textbf{checklists} of key reasoning points, covering functional intent, dependency relationships, and potential code effects—to support consistent LLM-as-a-Judge evaluation. 
This approach ensures that each instance contains both high-quality queries and verifiable reasoning anchors, enhancing its diagnostic value in assessing model understanding.

\item \textbf {{Reverse Masking.}} For tasks related to deployment and test generation, we employed a reverse construction strategy starting from verified “golden” artifacts and introducing controlled perturbations. 
For \textbf{\textit{Configuration \& Deployment}}, we randomly removed or replaced dependency packages in the \texttt{Dockerfile}, generating cases that may produce \textbf{Failed Docker Files}. Only those triggering reproducible build failures or functional inconsistencies (P2F) were retained, enabling evaluation of models’ ability to detect and fix deployment issues.
For \textbf{\textit{Test Case Generation}}, we selected PRs with more than five new test functions and formulated prompts using the corresponding \texttt{Code Patch} and \texttt{Test Patch}. GPT-5 generated evaluation queries such as \textit{``Generate unit tests to verify the correctness of the following patch.''} 
Instances that resulted in \textbf{Incomplete Tests} were retained to assess models’ capability to produce complete and correct test suites. Correctness and coverage metrics served as quantitative evaluation criteria.
\item \textbf {{Heuristic Filtering.}}
For patch-based tasks such as \textit{Performance Optimization}, \textit{Refactoring}, \textit{Feature Enhancement}, \textit{Feature Implementation}, and \textit{Bug Fixing}, we directly leveraged real-world PRs and applied targeted filtering rules to identify representative examples. 
Specifically, PRs that passed all unit tests both before and after patch application but exhibited runtime performance improvements exceeding 30\% were labeled as \textbf{\textit{Performance Optimization}} seeds. 
For \textbf{\textit{Refactoring}} tasks, we selected PRs that introduced substantial structural or readability improvements (e.g., function decomposition, code abstraction, or naming consistency) without altering external functionality.  
We then used GPT-5 to verify whether the changes explicitly addressed performance concerns described in the associated \texttt{Issue}. 
For the remaining three task types, classification was guided by patch intent and behavioral context:  
(1) \textbf{\textit{Feature Implementation}} instances corresponded to cases introducing entirely new modules or functionalities; 
(2) \textbf{\textit{Feature Enhancement}} focused on improvements or extensions to existing components; and 
(3) \textbf{\textit{Bug Fixing}} captured patches that directly resolved error logs, exceptions, or failing tests. 
Each identified instance was validated for logical consistency, build reproducibility, and semantic clarity to ensure task fidelity.
\end{itemize}


\subsubsection{Step 5: Data Validation}
To ensure both diversity and quality in the final benchmark, we applied a structured sampling and validation process designed to balance task coverage, control instance difficulty, and guarantee overall dataset reliability.

\begin{itemize}[left=10pt]
\item \textbf{{Difficulty Filtering.}} Each candidate instance was first evaluated based on the number of modified files, the number of changed lines, and additional signals derived from multiple model inferences.  This screening process ensures that all retained samples exhibit moderate and meaningful problem complexity, making them suitable for rigorous model evaluation.
\item \textbf{{Task-balanced Sampling.}} Balanced sampling was performed to maintain diversity across task and scenario dimensions. Sampling weights were further adjusted to reflect realistic distributions across 10 major programming languages, aligning with real-world open-source practices.
\item \textbf{{Manual Verification.}} All sampled instances underwent expert validation to confirm executability, correctness, and semantic consistency between commits, queries, Docker images, and corresponding test cases. Only verified instances were retained in the benchmark. 
\end{itemize}

As a result, we constructed the comprehensive benchmark \textbf{SWE-Compass}, which contains \textbf{2,000 high-quality instances}, well-balanced across task categories, programming scenarios, and languages, providing a rigorous and representative evaluation framework for assessing the capabilities of large language models in real-world software engineering tasks.

\subsection{Evaluation Metrics}

For each type of task, we select appropriate evaluation metrics to measure the model's performance. These include:

\begin{enumerate}
    \item \textbf{Pass@1}: The fraction of resolved samples achieved under a single attempt with fixed decoding and resource budgets.
    \item \textbf{Performance Optimization Score}: A binary indicator (0/1). The score is 1 if the model’s optimized code passes a single test and the time spent on execution is less than 80\% of the time taken by the unoptimized code; otherwise, the score is 0.
    \item \textbf{Line Coverage}: This metric evaluates the extent to which the program code has been executed during test case execution. The formula for calculating line coverage is:
    \[
    \text{Line Coverage} = \frac{\text{Number of Executed Code Lines}}{\text{Total Number of Code Lines}} \times 100\%
    \]
    \item \textbf{LLM-As-A-Judge Score}: Following~\citep{zheng2023judgingllmasajudgemtbenchchatbot, zhang2025codecriticbench, zhang2025artifactsbench, li2025relook}), we use a large language model (LLM) to review the model output according to a checklist; the final score is the proportion of checkpoints passed by the model output. 
\end{enumerate}

For specific tasks, the following metrics are used.
For \textbf{Feature Implementation, Feature Enhancement, Bug Fixing, and Refactoring}, \textbf{Pass@1} is used to measure the model's performance. For \textbf{Performance Optimization}, the \textbf{Performance Optimization Score} is used to evaluate the model’s performance. For \textbf{Test Case Generation}, we employ \textbf{Line Coverage} to assess the quality of the test cases generated by the model. In our implementation, we use pytest ~\citep{pajankar2017pytest} to compute line coverage for Python. For TypeScript and JavaScript, we use C8~\citep{bcoe2025c8} to calculate line coverage. For \textbf{Code Understanding}, we use the \textbf{LLM-As-A-Judge Score} to evaluate the accuracy of the model's understanding of the code. The specific prompt can be found in the Appendix \ref{app:CU_Judge_Prompt}.

%% file: sec/4_experiments.tex
\section{Experiments}
\label{sec:experiments}

\subsection{Evaluated LLMs and Frameworks}
\paragraph{Benchmarks and Tracks}
We evaluate \benchmark{} under two tracks (Executable and Non-executable). By default, we aggregate distribution-aligned over Task Type $\times$ Programming Scenario $\times$ Language following \S\ref{sec:method} with fixed seeds. Construction scale and composition are in \S\ref{sec:method} (Table~\ref{tab:benchmark_comparison}).

\paragraph{Frameworks}
We evaluate two offline agent workflows with identical executors: \textbf{SWE-Agent} (hardened edit--diff--execute loop) and \textbf{Claude Code} (sandboxed, editor-centric with parallel tool calls). Both use containerized, network-disabled toolchains with standardized build/test commands and execution hardening for reproducibility; complete workflow notes (including the parallel tool-call prompt) and the command matrix are in Appendix~\ref{app:exec_details}.


\paragraph{Environment, Budgets, and Metrics}
All evaluations run in fixed offline containers with unified budgets and executors; networking is disabled, and retries are not used. We adopt a single-attempt setting with standard decoding and turn/time limits, and evaluate using the task-type–aligned decision rules and metrics defined in \S\ref{sec:method}. Exact configuration (timeouts, hardware, container versions, context windows, caches) and any method-specific deviations are provided in Appendix~\ref{app:exec_details}.

\paragraph{LLMs}
We evaluate \textbf{10} models under a unified leaderboard (no reasoning/non-reasoning split): Claude-Sonnet-4-20250514, Qwen3-Coder-480B-A35B-Instruct, Qwen3-Coder-30B-A3B-Instruct, Qwen3-235B-A22B-Instruct-2507, Kimi-K2-Instruct-0905, Gemini-2.5-Pro, Gemini-2.5-Flash, GPT-4.1-2025-04-14, DeepSeek-V3-0324, and SWE-agent-LM-32B. Model API pages and open-source deployment links are provided in Appendix~\ref{app:model_links}.

\subsection{Experimental Results}
\subsubsection{Main Results}
\label{sec:main_leaderboard}

\begin{table}[t]
\centering
\small
\caption{Main results by task types on \benchmark{}.
AVG is the macro-average across task types. \textbf{Abbreviations}: FI=Feature Implementation; FE=Feature Enhancement; BF=Bug Fixing; RF=Refactoring; PO=Performance Optimization; CU=Code Understanding; TG=Test Case Generation; CD=Configuration \& Deployment.}
\label{tab:main_leaderboard}
\begin{tabular}{l|lllllllll}
\toprule
\multirow{2}{*}{\textbf{Models}} & \multicolumn{9}{c}{\textbf{Scores on Different Task Types}} \\
\cmidrule(lr){2-10}
 & \textbf{FI} & \textbf{FE} & \textbf{BF} & \textbf{RF} & \textbf{PO} & \textbf{CU} & \textbf{TG} & \textbf{CD} & \textbf{AVG} \\
\midrule
\multicolumn{10}{c}{\textit{SWE-Agent}}\\
\midrule
Claude-Sonnet-4 & 16.5 & \textbf{32.1} & \textbf{24.5} & \textbf{28.2} & \textbf{32.8} & \textbf{56.8} & 25.0 & \underline{51.0} & \underline{31.8} \\
Qwen3-Coder-480B-Instruct & \underline{17.9} & 25.4 & 20.3 & 24.1 & 26.4 & 42.1 & 23.4 & 49.5 & 27.2 \\
Kimi-K2-Instruct & 13.7 & 21.0 & 12.7 & 22.4 & 26.4 & 48.2 & 12.4 & 42.5 & 22.7 \\
Gemini-2.5-Pro & 11.8 & 17.4 & 13.1 & 14.9 & \underline{28.9} & 51.8 & 16.8 & 40.5 & 22.4 \\
GPT-4.1 & 14.2 & 16.1 & 13.5 & 19.5 & 21.4 & 43.0 & 18.5 & 39.0 & 21.4 \\
Qwen3-Coder-30B-Instruct & 12.3 & 21.9 & 12.4 & 23.6 & 24.4 & 38.4 & 10.8 & 37.0 & 20.7 \\
Qwen3-235B-A22B-Instruct & 10.4 & 19.2 & 12.2 & 19.5 & 17.9 & 41.1 & 18.5 & 23.0 & 18.8 \\
Gemini-2.5-Flash & 11.3 & 13.4 & 10.1 & 13.2 & 22.4 & 47.7 & 12.6 & 32.0 & 18.5 \\
Deepseek-V3 & 9.4 & 14.3 & 7.7 & 17.2 & 16.9 & 29.2 & 12.8 & 40.0 & 16.5 \\
SWE-agent-LM-32B & 10.4 & 11.6 & 9.0 & 15.5 & 13.9 & 17.9 & 14.7 & 35.0 & 14.7 \\
\midrule
\multicolumn{10}{c}{\textit{Claude Code}}\\
\midrule
Claude-Sonnet-4 & \textbf{21.2} & \underline{31.7} & \underline{24.0} & \underline{25.9} & 24.9 & \underline{56.5} & \textbf{28.4} & \textbf{65.5} & \textbf{32.9} \\
Qwen3-Coder-480B-Instruct & 11.8 & 18.3 & 14.4 & 12.1 & 22.9 & 42.8 & \underline{27.3} & 38.5 & 21.9 \\
Qwen3-Coder-30B-Instruct & \underline{17.9} & 23.7 & 12.2 & 21.3 & 15.4 & 36.2 & 21.8 & 28.0 & 21.6 \\
Qwen3-235B-A22B-Instruct & 6.1 & 12.5 & 9.0 & 13.8 & 15.4 & 36.0 & 13.9 & 21.0 & 14.7 \\
Deepseek-V3 & 4.7 & 8.0 & 6.6 & 6.3 & 6.0 & 22.3 & 11.1 & 19.0 & 9.8 \\
\bottomrule
\end{tabular}
\end{table}

Table~\ref{tab:main_leaderboard} reports Pass@1 by task type. Claude-Sonnet-4-20250514 ranks first under both workflows (32.9\% with Claude Code; 31.8\% with SWE-Agent). Scores largely cluster in the low-to-mid 20s (overall range roughly 10--33\%). Contrary to a monotonic advantage, the two workflows are complementary: among the five overlapping models, only two achieve higher AVG with Claude Code, whereas three obtain higher AVG with SWE-Agent. Among open-weight systems, Qwen3-Coder-480B-A35B-Instruct reaches 27.2\% with SWE-Agent and 21.9\% with Claude Code, still below the best proprietary model.
The findings across task types and workflows are as follows:

\paragraph{Findings by task type.}
A consistent but nuanced hierarchy emerges (Table~\ref{tab:main_leaderboard}). Code Understanding (CU) is among the strongest categories across models. Configuration \& Deployment (CD) can be high for some systems (e.g., Claude-Sonnet-4) but exhibits sizable cross-model variance, so it is not uniformly easy. Feature Enhancement (FE) and Refactoring (RF)—occupy a middle tier. Feature Implementation (FI) and Bug Fixing (BF) are harder, reflecting localization and integration challenges. Test Case Generation (TG) and Performance Optimization (PO) remain challenging, but not to single-digit averages; results typically fall in the mid-teens to mid-20s depending on the model. Method-wise, SWE-Agent tends to be stronger on BF and parts of FI that benefit from iterative localization, whereas Claude Code shows advantages on TG and some CD cases with more deterministic signals; CU is broadly comparable across the two.

\paragraph{Framework comparison.}
Across most settings, the two agents exhibit complementary strengths.
Mechanistically, SWE-Agent’s edit–diff–execute loop favors investigative, multi-file tasks that reward iterative localization, at the cost of higher timeout exposure; Claude Code’s sandboxed, editor-centric workflow yields strong performance on well-scoped, deterministic tasks (e.g., CD, CU, TG), benefitting from lower tool overhead. We also observe a trade-off with interaction efficiency: improvements in Eval score often coincide with higher average interaction turns (cf. Figure~\ref{fig:pass1_by_language_agent} and Figure~\ref{fig:turns_by_language_distribution}), with diminishing returns beyond moderate turn counts, suggesting that future gains require better localization and hypothesis pruning rather than simply more exploration.

\subsubsection{Further Analysis}
\label{sec:dist_slices}

\begin{figure}[t]
    \centering
    \includegraphics[width=0.95\textwidth]{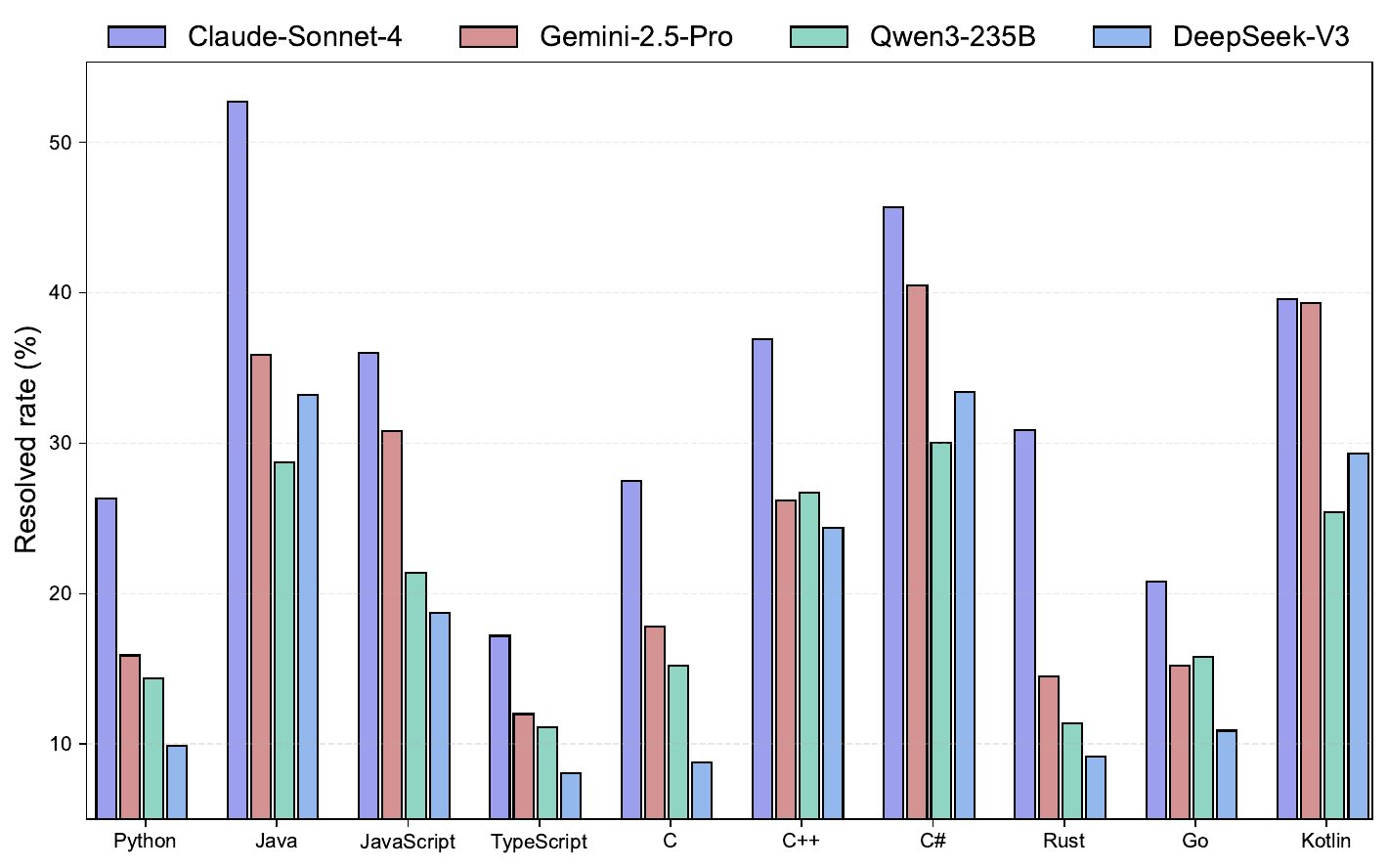}
    \caption{Comparison of Pass@1 (\%) across the top programming languages for SWE-Agent. Bars represent Pass@1; languages are ordered by overall Pass@1. This plot highlights whether improvements are concentrated in specific languages.}
    \label{fig:pass1_by_language_agent}
\end{figure}

\begin{figure}[t]
    \centering
    \includegraphics[width=\textwidth]{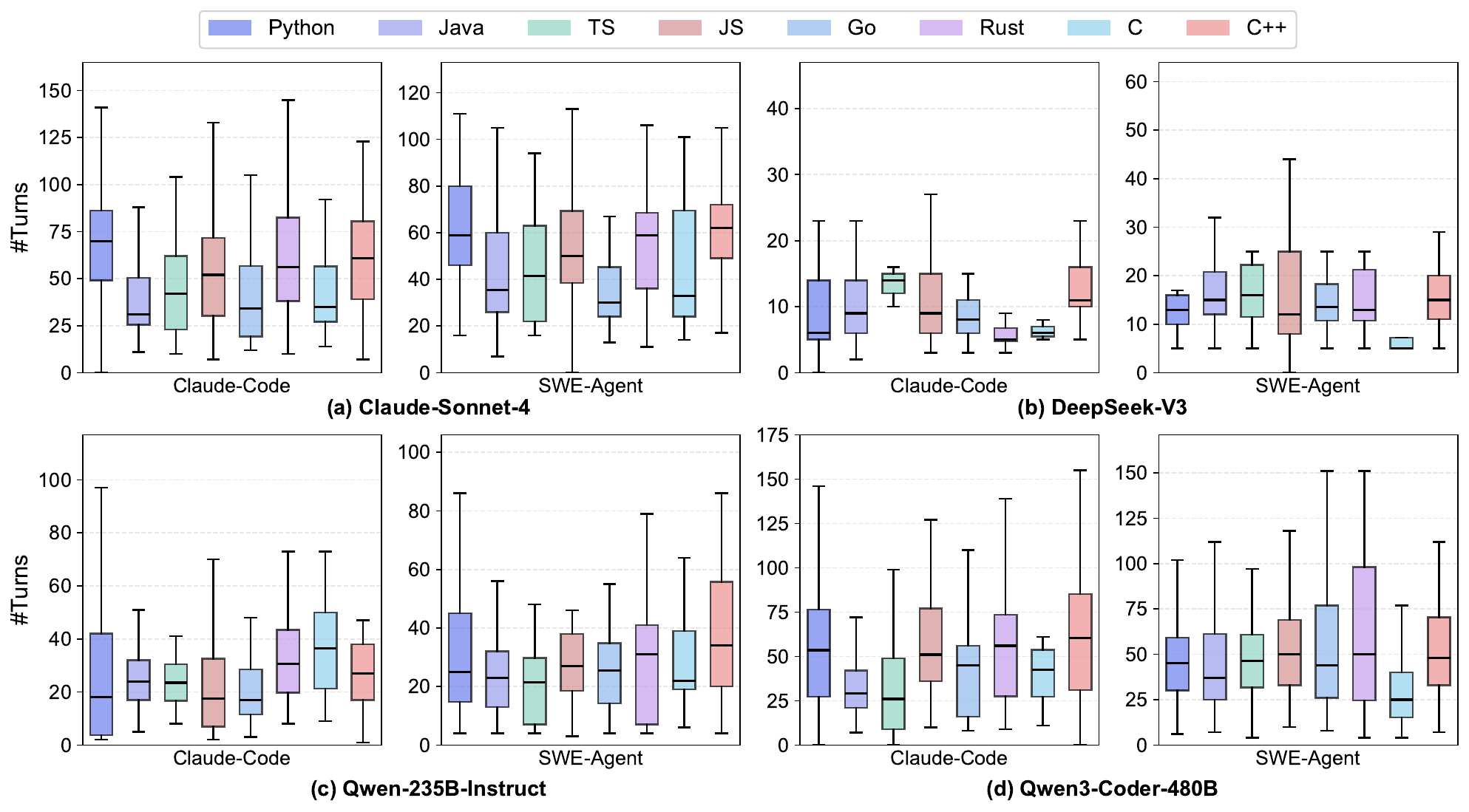}
    \caption{Distribution of interaction turns required per language for selected models to reveal trade-offs between effort (turns) and success. This highlights whether models achieve high Pass@1 by spending more turns on particular languages.}
    \label{fig:turns_by_language_distribution}
\end{figure}

\paragraph{Language-level observations.}

\begin{figure}[!ht]
    \centering
    \includegraphics[width=\textwidth]{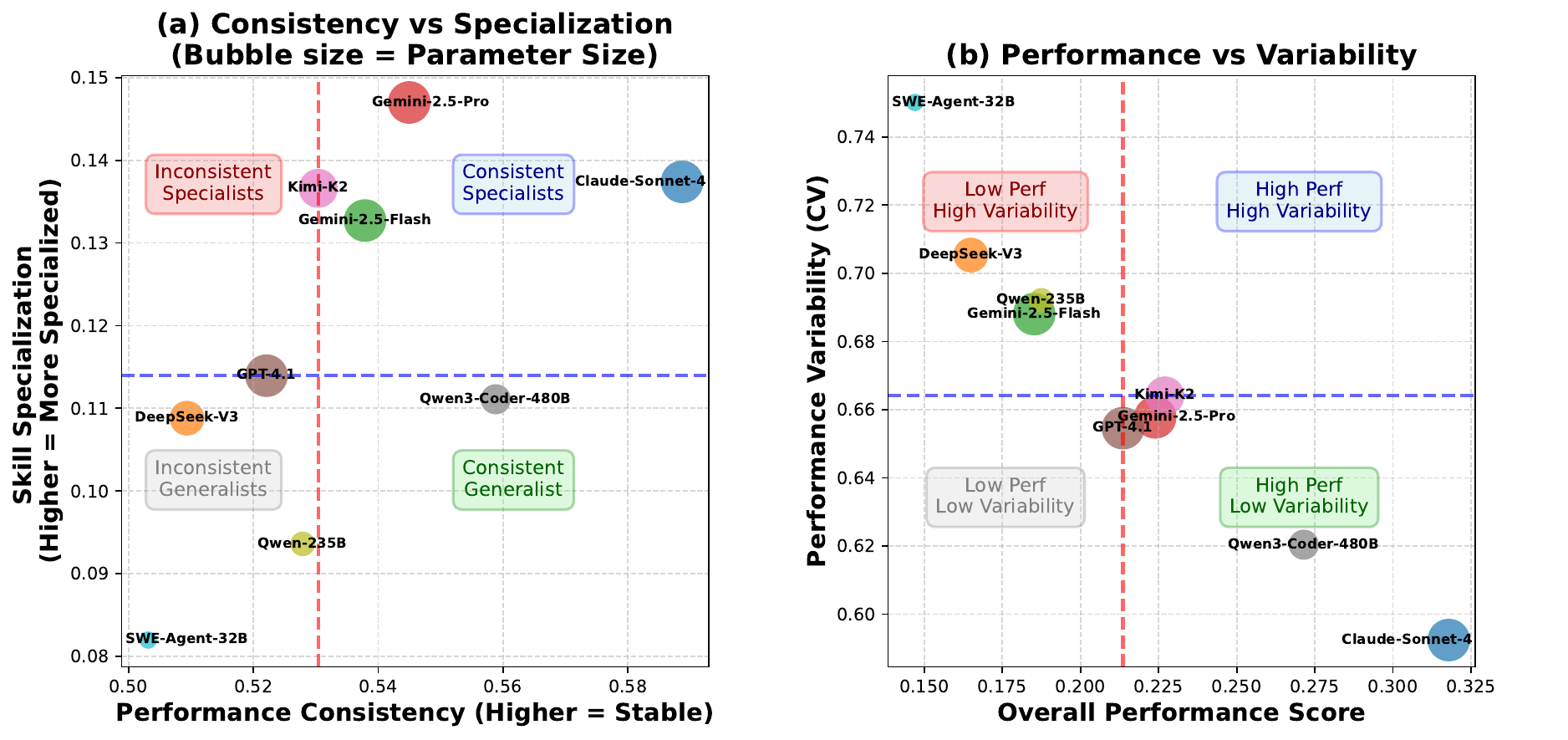}
    \caption{(Two panels)(a) This plot illustrates the relationship between skill specialization and performance consistency. Skill specialization is measured by the variability in performance across different tasks/languages, with higher values indicating stronger performance in specific tasks/languages compared to others. Performance consistency, on the other hand, reflects how stable the performance is across different difficulty levels, with higher values indicating more consistent performance over time. (b) The graph shows the relationship between performance variability and overall performance. Performance variability measures how much performance fluctuates across different tasks/languages, with higher values indicating greater inconsistency. Overall performance is calculated as the average score across tasks/languages, with higher values indicating better overall performance.}
    \label{fig:consistency_corr_variance}
\end{figure}

Figure~\ref{fig:pass1_by_language_agent}
and Appendix~Table~\ref{tab:lang_top10} indicate a consistent cross-language stratification across models and agents. JVM ecosystems and JavaScript tend to score higher (Java/Kotlin/JavaScript), while TypeScript is notably lower; systems languages (C/C++/Rust/Go) are harder; Python appears mid-tier overall, partly reflecting dataset selection effects—open-source benchmarks over-index on difficult Python bug-fixing cases (Appendix~\ref{opensource_dis_limit}. For Claude-Sonnet-4, Claude Code shows gains on Java/JavaScript, but this is not universal across models; in C\#, C/C++/Rust/Go, SWE-Agent often matches or outperforms. These patterns suggest performance is governed more by tooling determinism and diagnosability than raw coding difficulty; prioritize repository-level localization and environment hardening for systems/Python stacks, and hypothesis pruning for deterministic JVM/JS pipelines. See Figure~\ref{fig:consistency_corr_variance} for a visualization.

\paragraph{Interaction turns vs. success (by language).}
Figure~\ref{fig:turns_by_language_distribution} shows per-language turn distributions with Pass@1 overlays. Deterministic ecosystems (Java/Kotlin/JavaScript/C\#) have lower medians and tighter IQRs under Claude Code, while achieving similar or higher Pass@1—gains come from reliable signals rather than more turns. Systems languages (C/C++/Rust/Go) exhibit heavier tails, especially for SWE-Agent, with clear diminishing returns; Rust is most brittle. Python shows high variance: pinned environments converge quickly under Claude Code, while heterogeneity pushes SWE-Agent to many low-yield turns. Overall, prioritize repository-level localization for systems languages and environment hardening for Python; in JVM/JS, focus on sharper hypothesis pruning and parallel validation.

\paragraph{Consistency and specialization across languages.}
Figure~\ref{fig:consistency_corr_variance} summarizes whether strong aggregate results are broad-based or concentrated. In panel (a), aggregate Pass@1 and the per-language median Pass@1 show a clear visual positive trend, indicating that higher-ranked systems tend to improve more \emph{consistently} across languages rather than relying on a single language. Top systems cluster toward the right with higher consistency, whereas mid-tier models are more dispersed with lower medians. In panel (b), higher overall performance visually coincides with lower cross-language/task variability (coefficient of variation, CV), suggesting that stronger models are generally less variable. Together these observations support our earlier findings: (i) improvements at the top reflect broad gains in localization and execution reliability rather than narrow specialization; (ii) reducing cross-language variance is an effective lever for closing the gap, especially on systems languages that contribute disproportionately to variability; and (iii) evaluation protocols should report both central tendency and dispersion to avoid overstating gains driven by a subset of languages.

 Notably, Claude-Sonnet-4-20250514 lies at the far right with one of the lowest variabilities, reflecting broad cross-language gains.

\begin{table}[t]
\centering
\caption{Scores on different scenarios:  
Abbreviations: AD=Application Development; DE=Data Science \& Engineering; DS=Database Systems; ID=Infrastructure Development; ML=Machine Learning \& AI; SE=Security Engineering; SPD=Specialized Programming Domains; UI/UX=UI/UX Engineering; AVG=macro-average.}
\label{tab:scenario}
\small
\begin{tabular}{l|lllllllll}
\toprule
\multirow{2}{*}{\textbf{MODEL}} & \multicolumn{9}{c}{\textbf{Scores on Different Scenarios}} \\
\cmidrule(lr){2-10}
 & \textbf{AD} & \textbf{DE} & \textbf{DS} & \textbf{ID} & \textbf{ML} & \textbf{SE} & \textbf{SPD} & \textbf{UI/UX} & \textbf{AVG} \\
\midrule
\multicolumn{10}{c}{\textit{SWE-Agent}}\\
\midrule
Claude-Sonnet-4 & \underline{31.3} & \textbf{33.1} & \textbf{28.8} & \textbf{29.2} & \textbf{35.1} & \underline{31.4} & \underline{29.5} & \underline{38.5} & \underline{31.8} \\
Qwen3-Coder-480B-Instruct & 27.3 & 30.1 & 22.5 & 21.1 & 30.6 & 31.3 & 24.3 & 32.9 & 27.2 \\
Kimi-K2-Instruct & 22.2 & 24.6 & \underline{28.7} & 19.4 & 23.3 & 23.4 & 17.7 & 27.1 & 22.7 \\
Gemini-2.5-Pro & 21.5 & 23.1 & 24.9 & 17.7 & 24.5 & 25.1 & 21.2 & 24.8 & 22.4 \\
GPT-4.1 & 18.8 & 25.5 & 21.9 & 16.3 & 28.5 & 26.5 & 19.4 & 17.8 & 21.4 \\
Qwen3-Coder-30B-Instruct & 20.2 & 25.0 & 21.3 & 17.1 & 24.3 & 23.4 & 16.0 & 21.8 & 20.7 \\
Qwen3-235B-A22B-Instruct & 15.9 & 21.2 & 24.2 & 13.3 & 24.8 & 19.6 & 16.2 & 21.2 & 18.8 \\
Gemini-2.5-Flash & 14.1 & 21.6 & 28.1 & 13.5 & 22.3 & 15.6 & 17.6 & 22.2 & 18.5 \\
Deepseek-V3 & 13.9 & 17.2 & 18.1 & 12.6 & 22.1 & 18.9 & 14.9 & 19.0 & 16.5 \\

SWE-agent-LM-32B & 13.5 & 16.9 & 14.2 & 9.8 & 13.7 & 17.3 & 15.7 & 18.2 & 14.7 \\
\midrule
\multicolumn{10}{c}{\textit{Claude Code}}\\
\midrule
Claude-Sonnet-4 & \textbf{33.5} & \underline{32.9} & 27.8 & \underline{26.6} & \underline{33.9} & \textbf{37.3} & \textbf{29.9} & \textbf{44.7} & \textbf{32.9} \\
Qwen3-Coder-480B-Instruct & 21.1 & 20.4 & 23.1 & 17.0 & 23.9 & 23.2 & 22.1 & 28.8 & 21.9 \\
Qwen3-Coder-30B-Instruct & 21.2 & 23.3 & 21.6 & 15.3 & 29.2 & 25.3 & 18.3 & 26.8 & 21.6 \\
Qwen3-235B-A22B-Instruct & 12.3 & 17.7 & 19.5 & 14.5 & 14.2 & 19.4 & 8.8 & 13.8 & 14.7 \\
Deepseek-V3 & 7.2 & 14.4 & 10.3 & 9.1 & 11.3 & 11.1 & 7.3 & 9.3 & 9.8 \\
\bottomrule
\end{tabular}
\end{table}

\paragraph{Fine-grained scenario analysis.}
Table~\ref{tab:scenario} shows that scenario difficulty closely tracks tooling determinism and the locality of required edits. High-scoring categories such as UI/UX Engineering, Security Engineering, and Application Development combine mature frameworks with clear oracles and fast-running tests, where Claude Code’s editor-centric workflow converts stable feedback into higher Pass@1 with fewer turns. In contrast, Database Systems, Infrastructure Development, ML/AI, and Specialized Programming Domains involve multi-stage builds, cross-process dependencies, or non-deterministic outputs; here, SWE-Agent’s iterative localization is often more resilient but also more exposed to timeouts. The consistent average advantage of Claude Code over SWE-Agent across scenarios in our setting) is therefore concentrated in pipelines with reliable, low-variance signals. To close the remaining gaps, future systems should (i) enhance repository-level observability and reproducibility for complex stacks (minimal repro scripts, pinned environments, artifact isolation), and (ii) invest in hypothesis pruning and parallel verification for deterministic stacks, where the bottleneck is search efficiency rather than raw exploration budget.

\paragraph{Failure Mode Analysis.}

\begin{figure}[t]
    \centering
    \includegraphics[width=1.0\columnwidth]{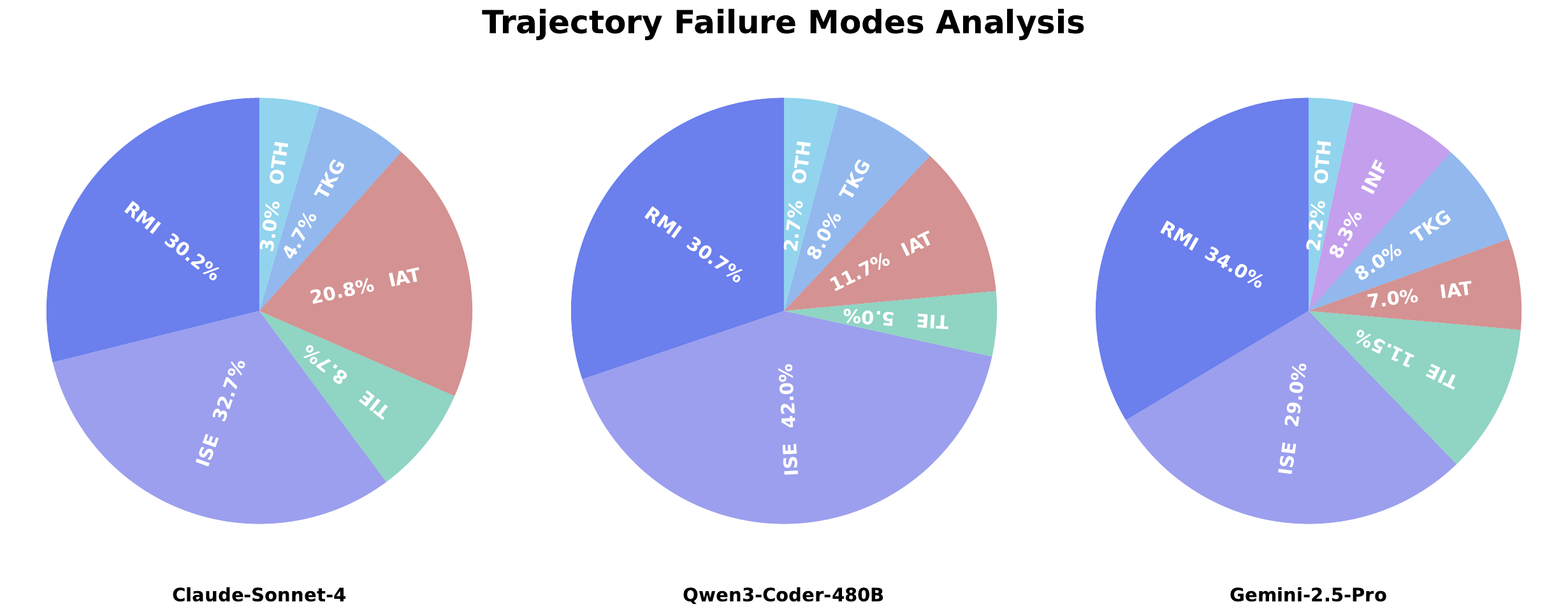}
    \caption{Distribution of trajectory failure modes on SWE-Compass. Abbreviations: RMI=Requirement Misinterpretation, ISE=Incomplete Solution \& Side Effects, TIE=Tool Invocation Error, IAT=Inadequate Testing, TKG=Technical Knowledge Gap, INF=Infinite Loop, OTH=Others.}
    \label{fig:failure_modes}
\end{figure}

To systematically understand the limitations of current coding agents, we perform a post-hoc failure analysis on \textbf{SWE-Agent} trajectories from our \benchmark{} benchmark. Following ~\cite{sweagent}—who report 87\% agreement between automated LLM judges and human experts—we adopt an \emph{LLM-as-Judge} protocol with \textbf{Claude-Sonnet-4} as the judge; the exact prompt is provided in Appendix~\ref{app:Trajectory_Failure_Analysis_Prompt}. We sample \textbf{600} failed trajectories \emph{per model} for three representative systems: \textbf{Claude-Sonnet-4}, \textbf{Qwen3-Coder-480B}, and \textbf{Gemini-2.5-Pro}.

Specifically, through manual inspection of submitted-but-failed trajectories, we develop a comprehensive six-category taxonomy capturing actual root causes:

\begin{enumerate}
    \item \textbf{Requirement Misinterpretation}: The agent failed to properly understand and locate the problem, including misidentifying affected files, misjudging severity, confusing problem types, or failing to identify root causes and understand dependencies, data flow, or system architecture.
    
    \item \textbf{Inadequate Testing}: The agent provided incomplete test coverage, missing edge cases, compatibility issues, performance impacts, integration scenarios, or multi-platform testing requirements.
    
    \item \textbf{Incomplete Solution \& Side Effects}: The agent provided an incomplete fix that only addressed symptoms rather than root causes, or introduced new issues, including regressions, security vulnerabilities, environment configuration errors, data corruption risks, or breaking changes to existing functionality.
    
    \item \textbf{Technical Knowledge Gap}: The agent demonstrated insufficient technical proficiency or violated domain-specific conventions, including lacking necessary knowledge in specialized domains (UI/frontend, security, accessibility, DevOps, performance, analytics) or incorrectly handling domain-specific issues (data processing, security implementations, UI/UX standards, API design, documentation synchronization).
    
    \item \textbf{Tool Invocation Error}: The agent encountered errors while using tools due to incorrect syntax, context overflow from file operations, or parse/analysis tool failures.
    
    \item \textbf{Infinite Loop}: The agent got stuck in loops without convergence, including repeated attempts at the same solution, oscillating between decisions, or endlessly reading files without making progress.
\end{enumerate}

Note that \textbf{OTH (Other)} denotes rare cases, which are not covered by the above taxonomy (e.g., corrupted artifacts or external executor glitches). The figure caption lists all abbreviations for completeness. As shown in Figure~\ref{fig:failure_modes}, we report the per-model distribution of failure modes on \benchmark{}. Based on \textbf{600} error traces per model, we draw the following conclusions: (1) \textbf{Shared bottlenecks in comprehension and implementation}. All models exhibit high error rates in \textbf{Requirement Misinterpretation} (30--34\%) and \textbf{Incomplete Solution \& Side Effects} (29--42\%), together accounting for \(>60\%\) of failures. By contrast, \textbf{Technical Knowledge Gap} is consistently low (5--8\%), suggesting the core limitations lie in requirement grounding and holistic solution design rather than basic coding proficiency. (2) \textbf{Distinct model characteristics}. \textit{Claude-Sonnet-4} is the most balanced, showing the lowest \textbf{Technical Knowledge Gap} (4.7\%) but room to improve on \textbf{Inadequate Testing} (20.8\%). \textit{Qwen3-Coder-480B} has the highest \textbf{Incomplete Solution \& Side Effects} rate (42\%, vs. Claude-Sonnet-4's 32.7\%), revealing weaknesses in end-to-end design. \textit{Gemini-2.5-Pro} shows the highest \textbf{Requirement Misinterpretation} (34\%) and a notable \textbf{Infinite Loop} issue (8.3\%), posing reliability risks in production. 

\section{Conclusion}
We introduced SWE-Compass, a unified benchmark that enables systematic evaluation of large language models across diverse software engineering tasks, scenarios, and languages. By integrating 2,000 verified instances derived from real-world GitHub repositories with reproducible execution environments, SWE-Compass provides comprehensive coverage of the software development lifecycle. Our large-scale experiments with ten state-of-the-art LLMs under two agentic frameworks reveal consistent hierarchies of task difficulty, language-specific variability, and dominant failure modes rooted in requirement misinterpretation and incomplete solutions. These findings highlight that future progress in automated software engineering depends less on isolated code generation improvements and more on enhancing requirement grounding, environment reliability, and reasoning consistency. SWE-Compass offers a rigorous, scalable, and reproducible foundation for advancing the next generation of robust, general-purpose coding agents.

\section{Future Works}

We see several directions to extend \benchmark{} and strengthen the community’s ability to measure and drive progress:
\begin{itemize}[left=10pt]
    \item \textbf{Scale and coverage}. Expand the dataset size, languages (e.g., mobile stacks and diverse SQL dialects), and repository types (monorepos, polyglot services), while maintaining distribution alignment across task, scenario, language, and difficulty.
    \item \textbf{Harder long-context settings}. Introduce multi-module, cross-process, and build-pipeline tasks that stress architectural coherence, multi-file reasoning, and cross-session memory under strict executability.
    \item \textbf{Metrics and protocols}. Enrich task-type–aligned metrics with long-context diagnostics (e.g., consistency and variance reporting), stabilize timing/coverage signals, and unify cost/efficiency reporting (turns, wall-clock, tool invocations) under fixed budgets.
    \item \textbf{Evaluation tracks}. Explore safe online or incremental tracks with evolving repositories and dependency drift, paired with sandboxing, artifact isolation, and replayable logs for fair comparison over time.
    \item \textbf{Human-in-the-loop calibration}. Establish human adjudication subsets and reliability audits for LLM-as-Judge, improving rubric calibration and model–human agreement on non-executable tasks.
    \item \textbf{Reproducibility, safety, and accessibility}. Continue releasing containers, minimal repro scripts, and a lightweight subset with stable seeds; strengthen privacy/safety filtering and provide clearer documentation to lower the barrier to participation.
\end{itemize}

%% file: sec/appendix.tex
\appendix
\section{Appendix}

\subsection{Judge Prompt for Code Understanding Task}
\label{app:CU_Judge_Prompt}
\begin{promptbox}{Judge Prompt for Code Understanding Task}

\begin{verbatim}

Evaluate if the answer satisfies the question requirements using a natural 
language explanation.

QUESTION:
{question_text}

REQUIREMENTS (checklist items):
{chr(10).join(checklist_text)}
{patch_section}
ANSWER:
{truncated_answer}

EVALUATION RULES:
1. Answer MUST use clear English explanations, NOT just code diffs, 
or other types of content
2. Only give 1.0 when ALL checklist items are thoroughly satisfied 
with clear explanations.
3. Score = (satisfied items) / (total items)
4. Penalize: code diffs without explanation, vague statements, 
wrong info. Give 0.0 when the answer is just code diffs or 
completely wrong.

JSON response:
{{
  "reasoning": "Brief explanation of which items satisfied/unsatisfied and why",
  "score": ,
  "satisfied_items": ["item_id1", ...]
}}

\end{verbatim}
\end{promptbox}

\subsection{Claude Code: Parallel Tool-Calls System Prompt}
\label{app:claude_parallel_prompt}
\noindent The following prompt is appended via SDK to encourage parallel tool invocations when operations are independent.

\begin{promptbox}{Claude Code: Parallel Tool-Calls System Prompt}

\begin{verbatim}

<use_parallel_tool_calls>
For maximum efficiency, whenever you perform multiple independent operations,
invoke all relevant tools simultaneously rather than sequentially. Prioritize
calling tools in parallel whenever possible. For example, when reading 3 files,
run 3 tool calls in parallel to read all 3 files into context at the same time.
When running multiple read-only commands like `ls` or `list_dir`, always run all
of the commands in parallel. Err on the side of maximizing parallel tool calls
rather than running too many tools sequentially.
</use_parallel_tool_calls>
\end{verbatim}
\end{promptbox}




\subsection{Trajectory Failure Analysis Prompt}
\label{app:Trajectory_Failure_Analysis_Prompt}
\begin{promptbox}{Trajectory Failure Analysis Prompt}

\textbf{System Prompt:}

\begin{verbatim}
You are an expert software engineer analyzing why a software engineering 
agent failed to resolve an issue.

AVAILABLE AGENT ACTIONS:

---- BEGIN FUNCTION #1: bash ----
Description: Execute a bash command in the terminal.
* Can generate very large outputs when listing files (ls, find, grep)
* Output contributes directly to context window usage
* Commands like 'find /repo -name "*.py"' can list thousands of files
* Large outputs can quickly fill the context window

Parameters:
  (1) command (string, required): The bash command to execute. Can be 
      empty to view additional logs when previous exit code is `-1`. 
      Can be `ctrl+c` to interrupt the currently running process.
---- END FUNCTION #1 ----

---- BEGIN FUNCTION #2: submit ----
Description: Finish the interaction when the task is complete OR if the 
assistant cannot proceed further with the task.
* Used when agent thinks task is done (may be correct or incorrect solution)
* Also used when agent is stuck and cannot make progress
* No parameters are required for this function.
---- END FUNCTION #2 ----

---- BEGIN FUNCTION #3: str_replace_editor ----
Description: Custom editing tool for viewing, creating and editing files
* State is persistent across command calls and discussions with the user
* If `path` is a file, `view` displays the result of applying `cat -n`. 
  If `path` is a directory, `view` lists non-hidden files and directories 
  up to 2 levels deep
* Directory views can generate large outputs contributing to context usage
* The `create` command cannot be used if the specified `path` already 
  exists as a file
* If a `command` generates a long output, it will be truncated and marked 
  with `<response clipped>`
* The `undo_edit` command will revert the last edit made to the file at `path`

Notes for using the `str_replace` command:
* The `old_str` parameter should match EXACTLY one or more consecutive 
  lines from the original file. Be mindful of whitespaces!
* If the `old_str` parameter is not unique in the file, the replacement 
  will not be performed. Make sure to include enough context in `old_str` 
  to make it unique
* The `new_str` parameter should contain the edited lines that should 
  replace the `old_str`

Parameters:
  (1) command (string, required): The commands to run. Allowed options are: 
      `view`, `create`, `str_replace`, `insert`, `undo_edit`.
  (2) path (string, required): Absolute path to file or directory, 
      e.g. `/repo/file.py` or `/repo`.
  (3) file_text (string, optional): Required parameter of `create` command, 
      with the content of the file to be created.
  (4) old_str (string, optional): Required parameter of `str_replace` command 
      containing the string in `path` to replace.
  (5) new_str (string, optional): Optional parameter of `str_replace` command 
      containing the new string (if not given, no string will be added). 
      Required parameter of `insert` command containing the string to insert.
  (6) insert_line (integer, optional): Required parameter of `insert` command. 
      The `new_str` will be inserted AFTER the line `insert_line` of `path`.
  (7) view_range (array, optional): Optional parameter of `view` command when 
      `path` points to a file. If none is given, the full file is shown. 
      If provided, the file will be shown in the indicated line number range, 
      e.g. [11, 12] will show lines 11 and 12. Indexing at 1 to start. 
      Setting `[start_line, -1]` shows all lines from `start_line` to the 
      end of the file.
---- END FUNCTION #3 ----

---- BEGIN FUNCTION #4: file_viewer ----
Description: Interactive file viewer for opening and navigating files in 
the editor.
* open <path> [<line_number>]: Opens the file at path. If line_number is 
  provided, the view moves to include that line.
* goto <line_number>: Moves the window to show the specified line number.
* scroll_down: Moves the window down 100 lines.
* scroll_up: Moves the window up 100 lines.

Parameters:
  (1) command (string, required): One of `open`, `goto`, `scroll_down`, 
      `scroll_up`.
  (2) path_or_line (string/int, optional): For `open`, a path (and optional 
      line). For `goto`, a line number.
---- END FUNCTION #4 ----

---- BEGIN FUNCTION #5: search_tools ----
Description: Searching utilities for locating text or files within the 
workspace.
* search_file <search_term> [<file>]: Searches for search_term in file. 
  If file is not provided, searches the current open file.
* search_dir <search_term> [<dir>]: Searches for search_term in all files 
  in dir. If dir is not provided, searches in the current directory.
* find_file <file_name> [<dir>]: Finds all files with the given name in dir. 
  If dir is not provided, searches in the current directory.

Parameters:
  (1) subcommand (string, required): One of `search_file`, `search_dir`, 
      `find_file`.
  (2) arg1 (string, required): The search term or file name, depending on 
      subcommand.
  (3) arg2 (string, optional): Target file (for search_file) or directory 
      (for search_dir/find_file).
---- END FUNCTION #5 ----

---- BEGIN FUNCTION #6: edit_block ----
Description: Block editor for replacing ranges in the current open file 
and finalizing edits.
* edit <n>:<m> <replacement_text>: Replaces lines n through m (inclusive) 
  with the given text in the open file. Ensure indentation is correct.
* end_of_edit: Applies the pending changes. Python files are syntax-checked 
  after the edit; if an error is found, the edit is rejected.

Parameters:
  (1) command (string, required): `edit` or `end_of_edit`.
  (2) range_and_text (varies): For `edit`, a line range `n:m` and the 
      replacement text.
---- END FUNCTION #6 ----

---- BEGIN FUNCTION #7: create_file ----
Description: Creates and opens a new file with the given name.

Parameters:
  (1) filename (string, required): Absolute or workspace-relative path to 
      create. The file must not already exist.
---- END FUNCTION #7 ----

##PROBLEM STATEMENT##
{problem_statement}

##TRAJECTORY SUMMARY##
- Total steps: {total_steps}
- Final state: Failed (no successful patch generated / failed on some unit test)

##ANALYSIS INSTRUCTIONS##

**IMPORTANT: This trajectory FAILED in final evaluation. The agent likely 
believed it succeeded, but it was WRONG.**

The agent may have:
- Claimed the issue was resolved or fixed
- Written custom tests that passed
- Expressed high confidence in the solution
- Stated "the implementation is complete" or "all tests pass"
- Created demo scripts showing the fix "works"
- Manually verified outputs that looked correct

Despite these apparent indicators of success, the final evaluation proves 
the solution was INCORRECT. Therefore, ignore the agent's self-assessment 
and focus on identifying the actual flaws. Select ONE category below that 
best describes the actual flaw:
Requirement Misinterpretation: The agent failed to properly understand and 
locate the problem, including misidentifying affected files, misjudging 
severity, confusing problem types, or failing to identify root causes and 
understand dependencies, data flow, or system architecture.
Inadequate Testing: The agent provided incomplete test coverage, missing 
edge cases, compatibility issues, performance impacts, integration scenarios, 
or multi-platform testing requirements.
Incomplete Solution & Side Effects: The agent provided an incomplete fix 
that only addressed symptoms rather than root causes, or introduced new 
issues including regressions, security vulnerabilities, environment 
configuration errors, data corruption risks, or breaking changes to existing 
functionality.
Technical Knowledge Gap: The agent demonstrated insufficient technical 
proficiency or violated domain-specific conventions, including lacking 
necessary knowledge in specialized domains (UI/frontend, security, 
accessibility, DevOps, i18n, performance, analytics) or incorrectly handling 
domain-specific issues (data processing, security implementations, UI/UX 
standards, API design, documentation synchronization).
Tool Invocation Error: The agent encountered errors while using tools due 
to incorrect syntax, context overflow from file operations, or parse/analysis 
tool failures.
infinite_loop: The agent got stuck in loops without convergence, including 
repeated attempts at the same solution, oscillating between decisions, or 
endlessly reading files without making progress.

other: The agent failed to resolve the issue for reasons not covered by the 
above categories.

Do NOT invent or propose new categories. If none fits, use "other". Category 
must be all lowercase with underscores. Remember to write two new lines 
before the category.
\end{verbatim}

\textbf{User Prompt:}

\begin{verbatim}
##INSTANCE INFORMATION##
Instance ID: {instance_id}

##The complete trajectory of the interaction (to be analyzed)##
{traj_text}

##OUTPUT FORMAT##
You MUST provide your response in this exact format:
<description>
xxx
</description>

<category>
xxx
</category>

<error_actions>
If the Assistant gets stuck in a loop or encounters a tool_error error, 
indicate the incorrect action and parameters. If the Assistant misunderstands 
the question, set error_action="None".
</error_actions>
\end{verbatim}

\end{promptbox}

\subsection{Executor Details and Method-Specific Settings}
\label{app:exec_details}
\noindent Unless otherwise noted, all runs are strictly offline. Below we record method-specific configurations referenced in \S\ref{sec:experiments}:
\begin{itemize}[leftmargin=*]
  \item \textbf{SWE-Agent.} \emph{max turns} $=150$; per-tool step timeout $=600$ s; \texttt{parse\_function} set to \textit{function calling}; long observations truncated; compiled artifacts filtered via ``.gitignore''; language-specific build/test commands repaired for stability.
  \item \textbf{Claude Code.} \emph{max turns} $=150$; \texttt{permission\_mode} $=$ \textbf{bypassPermissions}; system prompt appended to encourage parallel tool calls (Appx.~\ref{app:claude_parallel_prompt}); may internally invoke a SubAgent; networking disabled.
\end{itemize}

\paragraph{Standardized offline build/test commands (per language).}
We standardize non-interactive commands to ensure reproducible builds and comparable feedback signals across languages:
\begin{itemize}[leftmargin=*]
  \item \textbf{Python}: \texttt{pytest -q}
  \item \textbf{JavaScript/TypeScript}: \texttt{npm ci \&\& npm test --run}
  \item \textbf{Java}: \texttt{mvn -B -DskipTests=false test}
  \item \textbf{Go}: \texttt{go test ./...}
  \item \textbf{Rust}: \texttt{cargo test --locked}
  \item \textbf{C/C++}: \texttt{cmake --build \&\& ctest -j1}
\end{itemize}

\paragraph{Execution hardening and navigation controls.}
To minimize offline flakiness and improve determinism, we apply:
\begin{itemize}[leftmargin=*]
  \item Pinned toolchains inside containers; pre-populated offline caches/proxies for \texttt{pip}, \texttt{npm}, \texttt{cargo}, Maven/Gradle, and Go modules.
  \item Truncation of long observations and logs; filtering of binaries and dev servers via ``.gitignore''.
  \item Normalization of EOL/encoding (LF, UTF-8); \texttt{git} \texttt{safe.directory} set; whitespace-tolerant patching.
  \item Repository navigation pruning by extensions; optional function/class extraction (read-only) to speed up localization.
  \item Budgets and quotas: max 150 turns; per-step 600 s; global job limits; auto-kill of long-lived processes.
  \item Language-specific repairs for brittle stacks (e.g., Java multi-module builds, Node lockfile drift, Rust workspaces, C/C++ out-of-tree builds).
\end{itemize}

\begin{table}[t]
\centering
\caption{Top-10 languages: Pass@1 (\%) per model. Columns are languages; rows are models grouped by agent.}
\label{tab:lang_top10}
\small
\resizebox{\textwidth}{!}{
\begin{tabular}{l|lllllllllll}
\toprule
\multirow{2}{*}{\textbf{MODEL}} & \multicolumn{11}{c}{\textbf{Scores on Different Languages}} \\
\cmidrule(lr){2-12}
 & \textbf{Python} & \textbf{Java} & \textbf{JavaScript} & \textbf{TypeScript} & \textbf{C} & \textbf{C++} & \textbf{C\#} & \textbf{Rust} & \textbf{Go} & \textbf{Kotlin} & \textbf{AVG} \\
\midrule
\multicolumn{12}{c}{\textit{SWE-Agent}}\\
\midrule
Claude-Sonnet-4 & \textbf{26.3} & 52.7 & 36.0 & 17.2 & \textbf{27.5} & \textbf{36.9} & 45.7 & \textbf{30.9} & 20.8 & 39.6 & 31.8 \\
Qwen3-Coder-480B-Instruct & 20.8 & 50.0 & 36.0 & 13.2 & 18.0 & 30.6 & 30.9 & 24.6 & 16.2 & 38.1 & 27.2 \\
Kimi-K2-Instruct & 13.2 & 43.0 & 24.4 & 13.7 & 15.8 & 33.4 & 42.2 & 15.2 & \textbf{21.9} & 28.2 & 22.7 \\
Gemini-2.5-Pro & 15.9 & 35.9 & 30.8 & 12.0 & 17.8 & 26.2 & 40.5 & 14.5 & 15.2 & 39.3 & 22.4 \\
GPT-4.1 & 17.2 & 31.5 & 25.0 & 10.0 & 16.2 & 28.3 & \textbf{46.8} & 15.8 & 12.7 & 34.7 & 21.4 \\
Qwen3-Coder-30B-Instruct & 13.5 & 32.1 & 26.3 & 13.7 & 13.6 & 28.7 & 28.4 & 14.9 & 15.5 & 37.8 & 20.7 \\
Qwen3-235B-A22B-Instruct & 14.4 & 28.7 & 21.4 & 11.1 & 15.2 & 26.7 & 30.0 & 11.4 & 15.8 & 25.4 & 18.8 \\
Gemini-2.5-Flash & 12.7 & 30.4 & 20.5 & 14.1 & 15.7 & 24.9 & 32.2 & 13.1 & 16.0 & 22.3 & 18.5 \\
Deepseek-V3 & 9.9 & 33.2 & 18.7 & 8.1 & 8.8 & 24.4 & 33.4 & 9.2 & 10.9 & 29.3 & 16.5 \\

SWE-agent-LM-32B & 11.1 & 30.5 & 20.8 & 7.2 & 4.7 & 17.0 & 20.9 & 6.5 & 11.7 & 18.7 & 14.7 \\
\midrule
\multicolumn{12}{c}{\textit{Claude Code}}\\
\midrule
Claude-Sonnet-4 & 23.5 & \textbf{60.0} & \textbf{46.3} & \textbf{18.5} & 24.5 & 34.5 & 43.2 & 28.2 & 18.2 & \textbf{51.9} & \textbf{32.9} \\
Qwen3-Coder-480B-Instruct & 17.7 & 39.4 & 30.3 & 15.2 & 14.9 & 21.1 & 27.4 & 20.7 & 13.2 & 21.4 & 21.9 \\
Qwen3-Coder-30B-Instruct & 16.0 & 41.1 & 28.8 & 13.6 & 16.2 & 22.1 & 23.7 & 11.5 & 20.7 & 33.3 & 21.6 \\
Qwen3-235B-A22B-Instruct & 10.4 & 18.7 & 21.5 & 9.0 & 12.3 & 14.5 & 20.3 & 11.8 & 12.4 & 32.1 & 14.7 \\
Deepseek-V3 & 7.1 & 14.4 & 15.0 & 4.1 & 5.7 & 11.7 & 16.0 & 4.7 & 8.0 & 20.6 & 9.8 \\
\bottomrule
\end{tabular}}
\end{table}

\begin{figure}[!ht]
    \centering
    \includegraphics[width=0.7\textwidth]{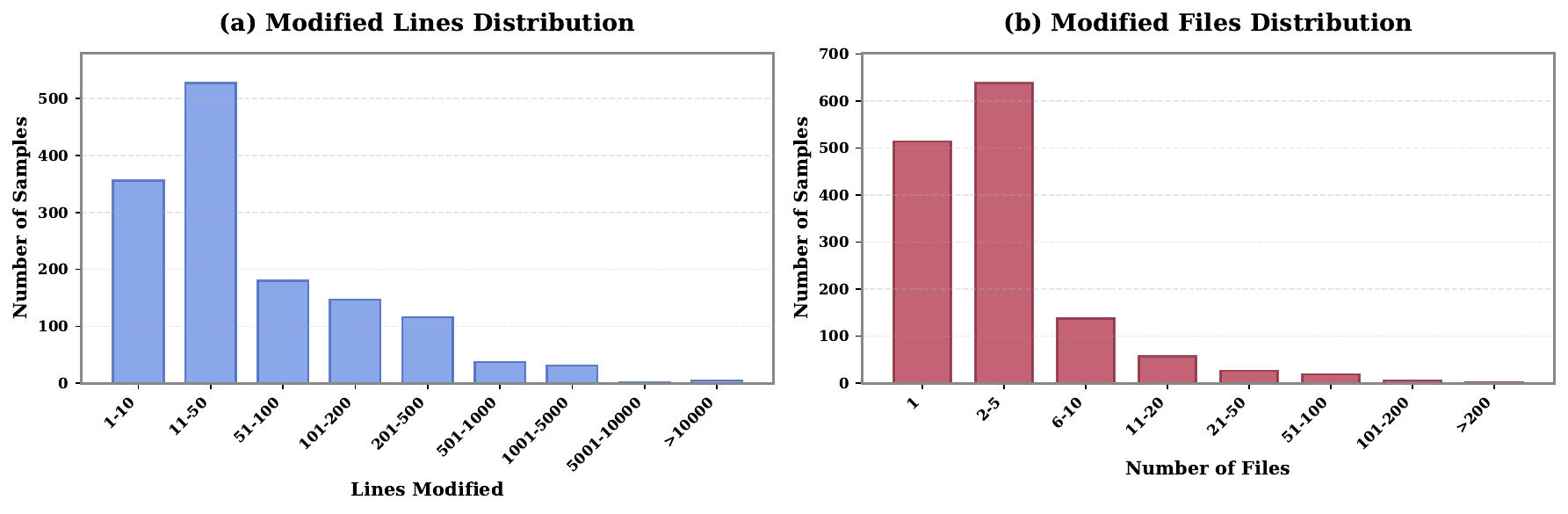}
    \caption{Distribution of modified files and lines involved in golden patches.}
    \label{fig:golden_patch_distribution}
\end{figure}

\subsection{Analysis of Open-Source Benchmark Distributions}
\label{opensource_dis_limit}
\begin{figure}[t]
  \centering
  \includegraphics[width=\linewidth]{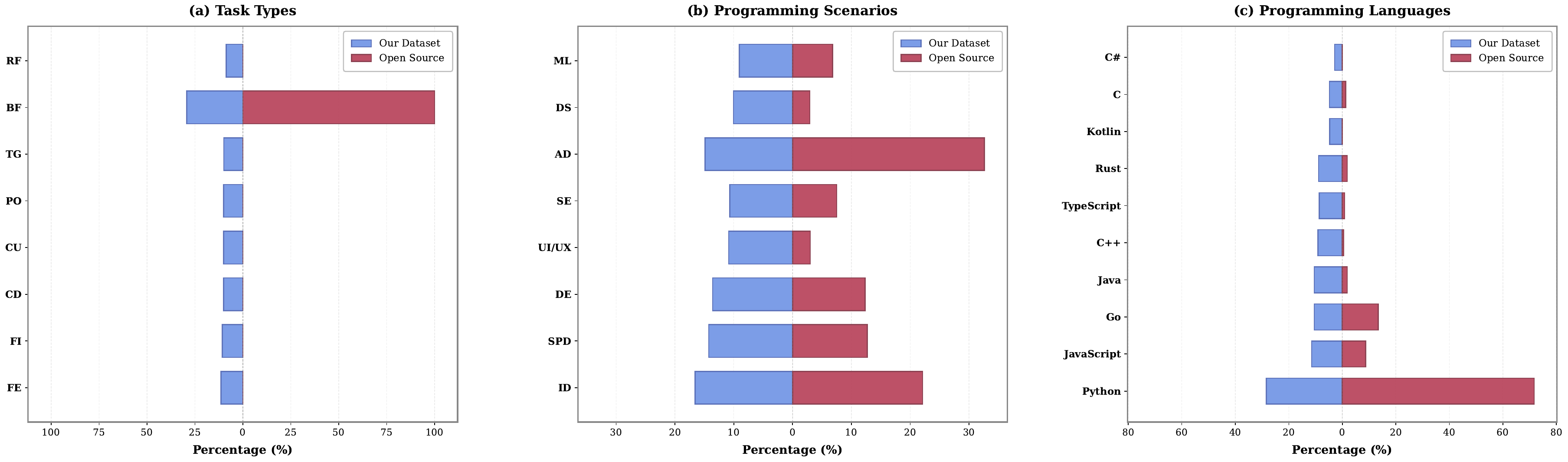}
  \caption{Distributions across task types, programming scenarios, and languages in Open-Source SWE Benchmarks and Github PR \& Issue. 
  \textbf{Abbreviations:} FE: Feature Enhancement, FI: Feature Implementation, CD: Configuration \& Deployment, CU: Code Understanding, PO: Performance Optimization, TG: Test Case Generation, BF: Bug Fixing, RF: Refactoring; ID: Infrastructure Development, SPD: Specialized Programming Domains, DE: Data Science \& Engineering, SE: Security Engineering, AD: Application Development, DS: Database Systems, ML: Machine Learning \& AI, UI/UX: UI/UX Engineering.}
  \label{fig:step2_collection_dis}
\end{figure}

We annotated several repository-level SWE benchmark datasets, including \textbf{SWE-bench-Verified} ($n=500$), \textbf{SWE-bench-Live} ($n=500$), \textbf{SWE-bench-Multilingual} ($n=300$), \textbf{SWE-bench-Pro} ($n=731$), and \textbf{SWE-rebench} ($n=449$), totaling \textbf{2,480} instances across multiple programming languages and scenarios.
Figure~\ref{fig:step2_collection_dis}(a) presents the distribution of these open-source datasets across task types, program scenarios, and languages. Through detailed analysis, we identified the following limitations:
\begin{itemize}[left=10pt]
\item \textbf{Incomplete Task Type Coverage.} Existing benchmarks are entirely focused on \textit{Bug Fixing} tasks, which comprise 100\% of all instances. In contrast, several important task types—including \textit{Feature Enhancement}, \textit{Feature Implementation}, \textit{Configuration \& Deployment}, \textit{Code Understanding}, \textit{Performance Optimization}, \textit{Test Case Generation}, and \textit{Refactoring}—are completely absent.

\item \textbf{Imbalanced Scenario Distribution.} A large portion of the data focuses on \textit{Application Development} (32.6\%), whereas other critical scenarios such as \textit{UI/UX Engineering} (3.0\%), \textit{Database Systems} (2.9\%), and \textit{Security Engineering} (7.5\%) receive significantly less coverage. Meanwhile, \textit{Infrastructure Development} accounts for 22.1\%.

\item \textbf{Severe Programming Language Imbalance.} The datasets are overwhelmingly dominated by \textit{Python} (71.7\%), with minimal coverage of other programming languages such as \textit{Go} (13.5\%), \textit{JavaScript} (8.7\%), and others combined accounting for less than 6\%.
\end{itemize}

\subsection{Model link list}
\label{app:model_links}

We evaluate our approach using a diverse set of state-of-the-art language models, including both closed-source and open-source models. The closed-source models include Claude-Sonnet-4-20250514~\citep{anthropic2025systemcard}, Gemini-2.5-Flash, Gemini-2.5-Pro~\citep{gemini2023gemini}, and GPT-4.1-2025-04-14~\citep{openai2024gpt4technicalreport}. For open-source models, we utilize Qwen3-Coder series~\citep{qwen3technicalreport,hui2024qwen2}, Kimi-K2-Instruct-0905~\citep{kimiteam2025kimik2openagentic}, Deepseek-V3-0324~\citep{liu2024deepseek}, and SWE-agent-LM-32B~\citep{yang2025swesmith}. The complete list of models with their official links is provided in Table~\ref{tab:model_list}.

\begin{table}[h]
\centering
\small 
\caption{Model List.}
\label{tab:model_list}
\begin{tabular}{l|l}
\toprule
\textbf{Model} & \textbf{Link} \\
\midrule
\multicolumn{2}{c}{\textit{Closed-Source Models}} \\
\midrule
Claude-Sonnet-4-20250514 & \url{https://www.anthropic.com/claude} \\
Gemini-2.5-Flash & \url{https://deepmind.google/technologies/gemini/flash/} \\
Gemini-2.5-Pro & \url{https://deepmind.google/technologies/gemini/pro/} \\
GPT-4.1-2025-04-14 & \url{https://openai.com/gpt-4} \\
\midrule
\multicolumn{2}{c}{\textit{Open-Source Models}} \\
\midrule
Qwen3-Coder-480B-A35B-Instruct & \url{https://huggingface.co/Qwen/Qwen3-Coder-480B-A35B-Instruct} \\
Qwen3-Coder-30B-A3B-Instruct & \url{https://huggingface.co/Qwen/Qwen3-Coder-30B-A3B-Instruct} \\
Qwen3-235B-A22B-Instruct-2507 & \url{https://huggingface.co/Qwen/Qwen3-235B-A22B-Instruct-2507} \\
Kimi-K2-Instruct-0905 & \url{https://huggingface.co/moonshotai/Kimi-K2-Instruct-0905} \\
Deepseek-V3-0324 & \url{https://huggingface.co/deepseek-ai/DeepSeek-V3} \\
SWE-agent-LM-32B & \url{https://huggingface.co/SWE-bench/SWE-agent-LM-32B} \\
\bottomrule
\end{tabular}
\end{table}